\documentclass[]{article}

\usepackage{amssymb,amsmath}
\usepackage{epic,eepic}
\usepackage{texdraw}
\usepackage{dsfont}
\usepackage{latexsym}

\usepackage{accents}
\DeclareMathAccent{\wtilde}{\mathord}{largesymbols}{"65}
\DeclareMathAccent{\what}{\mathord}{largesymbols}{"62}

\makeatletter
\def\wb{\accentset{{\cc@style\underline{\mskip10mu}}}}
\makeatother

\newcommand{\be}{\begin{eqnarray}}
\newcommand{\ee}{\end{eqnarray}}
\newcommand{\bdm}{\begin{displaymath}}
\newcommand{\edm}{\end{displaymath}}
\newtheorem{thm}{Proposition}
\newtheorem{ex}{Example}
\newtheorem{lm}{Lemma}
\newtheorem{deff}{Definition}
\newtheorem{rem}{Remark}
\begin{document}

\title{Affine linear and $\rm{D}_4$ symmetric lattice equations : symmetry analysis and reductions}

\author{A. Tongas, D. Tsoubelis and P. Xenitidis \\
Department of Mathematics, University of Patras, 265 00 Patras, Greece}

\maketitle

\begin{abstract}
We consider lattice equations on ${\mathds{Z}}^2$ which are autonomous, affine linear and possess the symmetries of the square. Some basic properties of equations of this type are derived, as well as a sufficient linearization condition and a conservation law. A systematic analysis of the Lie point and the generalized three- and five-point symmetries is presented. It leads to the generic form of the symmetry generators of all the equations in this class, which satisfy a certain non-degeneracy condition. Finally, symmetry reductions of certain lattice equations to discrete  analogues of the Painlev\'e equations are considered.
\end{abstract}

\section{Introduction}

The importance of symmetry based techniques applied to differential equations, especially to nonlinear ones, is well known. It can be argued that symmetry methods are the most effective ones for obtaining explicit solutions of complicated (systems of) nonlinear partial differential equations. In fact, the so-called group invariant solutions of such equations form a well known example of the results arrived at by these methods. It is also well known that, using its symmetries one can construct new interesting solutions of a given equation from much simpler ones. 

However, the notion of a symmetry group of transformations acting on the solution space does not have to be limited only to differential equations. It can be equally well applied to other types, such as the algebraic and difference ones. The latter arise in many diverse branches of mathematics and physics, such as discrete geometry, integrable systems, special functions and orthogonal polynomials, the study of exactly solvable models in statistical mechanics, crystal lattice theory and many others. The wide range of their applications shows that difference equations are of equal importance with their differential counterparts. As a result, symmetry methods have started being applied to the analysis of difference equations, as well. 

In fact, difference equations have already been studied by symmetry methods from various points of view, see e.g. \cite{byrnes, tp:Clark, tp:Dor3, tp:Hydon, tp:Levi3, tp:Levi1, tp:maeda, tp:Quisp1} and references therein. Symmetries of integrable partial difference equations first appeared as compatible constraints in the work of Nijhoff and Papageorgiou \cite{Pap2}. The motivation was a specific reduction of the discrete modified Korteweg-de Vries (KdV) equation to a discrete analog of Painlev\'e II, in the same spirit as the Painlev\'e II ordinary differential equation arises as a similarity reduction of the celebrated partial differential equation of KdV. Further examples of such compatible constraints for integrable partial difference equations were given in \cite{tp:ramani,tp:Nij3,frank,brew2,tp:TTP}.

In this paper, we present a systematic study on the symmetries and reductions of autonomous partial difference equations, which are not necessarily integrable. Specifically, we consider a quite large class of lattice equations defined on an elementary quadrilateral, which contains the integrable ones classified recently by Adler, Bobenko and Suris in \cite{ABS}. The members of this class are characterized by (i) affine linearity and (ii) ${\mathrm{D}}_4$-symmetry i.e. the symmetries of the square. 

We first prove that each equation in this class admits two, at least, three-point generalized symmetries. They are determined by a pair of polynomials arising from the equation's defining relation. For the generic case, we give the form of the generators of three- and five- point generalized symmetries, and a greatly simplified form of the corresponding determining equations. These results also extended to symmetry transformations which act on the lattice parameters appearing in the equation, as well. 

The previous symmetry analysis is then applied to the equations obtained in classification \cite{ABS}. The result of this part of our investigation is an exhaustive  list of the corresponding Lie point, three- and five-point generalized symmetry generators. As a final application, we consider specific symmetry reductions of the discrete potential KdV to ordinary difference equations, which represent discrete analogues of those of Painlev\'e. 

The paper is organized as follows. Section \ref{sec1} contains the necessary preliminaries on symmetries of difference equations and the notation that we use in the following sections. In Section \ref{sec3} we introduce the family of lattice equations under consideration and its characteristic properties.
The main results of the symmetry analysis are presented in the following four sections, where the general form of the corresponding symmetry generators is given explicitly. The Lie point symmetries are studied in Section \ref{lie}. The three- and five-point generalized symmetries are presented in Sections \ref{3point} and \ref{5point}, respectively. In Section \ref{param} we extend the previous considerations to symmetry transformations acting on the lattice parameters, as well.
In Section \ref{symabs} we present the symmetries of the equations of the classification \cite{ABS}, and Section \ref{red} deals with symmetry reductions. We conclude with Section \ref{concl}, where an overall evaluation of the results obtained in the main body of the paper is presented, along with various perspectives on the subject. In the Appendix, a detailed proof of the Proposition of Section \ref{3point} is given.

\section{Preliminaries on symmetries of difference equations}
\label{sec1}

A partial difference equation is a functional relation among the values of a function $u : \mathds{Z} \times \mathds{Z} \rightarrow \mathds{C}$ (or $\mathds{CP}$) at different points of the lattice, which in general involves the independent variables $n$, $m$ and the lattice spacings $\alpha$, $\beta$, as well, i.e. a relation of the form
\begin{equation}
\mathcal{E}\left(n,m,u(n,m),u(n+1,m),u(n,m+1),\ldots;\alpha,\beta\right)\, =\, 0 \,. \label{eq:eq}
\end{equation}

The analysis of partial difference equations is facilitated by the use of two translation operators on functions on ${\mathds{Z}}^2$, defined by 
$$\left(\mathcal{S}_n^{(k)} u\right)(n,m) = u(n+k,m)\,,\,\,\, \left(\mathcal{S}_m^{(k)} u\right)(n,m) = u(n,m+k)\,,\,\,\, {\mbox{where}}\, k \in \mathds{Z} \,.$$
It is also found useful to introduce the notation
\begin{equation}
u(n,m)\,=\,u_{(0,0)}\,,\,\,\,u_{(k,l)} \,=\,u(n+k,m+l)\,,\quad{\mbox{where}}\,\,k,\,l\,\in\,{\mathds{Z}}\,,
\end{equation}
for the values of the function $u$ and this will be adopted from now on.

Let $\mbox{G}$ be a one-parameter group of transformations acting on $\mathds{CP}$, the domain of the dependent variable $u_{(0,0)}$ of a lattice equation, i.e.
$${\mbox{G}}\,:\,u_{(0,0)}\,\rightarrow\,\Phi(n,m,u_{(0,0)};\varepsilon)\,,\quad \varepsilon \,\,\in\,\,{\mathds{C}}\,. $$
We denote by ${\rm{J}}^{(k)}$ the forward lattice jet space of order $k \in {\mathds{N}}$ with coordinates $(u_{(i,j)})$, where $i,j \in {\mathds{N}}$ and $i+j \le k$ . Similarly,
one can define the backward lattice jet space of order $k$, denoted by ${\rm{J}}^{(-k)}$, with coordinates $(u_{(-i,-j)})$, $i,j \in {\mathds{N}}$ and $i+j \le k$,
and in general the $k$-order lattice jet space ${\rm{J}}^{(k,-k)}$, with coordinates $\big(u_{(\pm i,\pm j)}\big)$, $i,j \in {\mathds{N}}$ and $i+j \le k$. The prolongation of the group action of $G$ on ${\rm{J}}^{(k)}$ is defined by
\begin{equation}
G^{(k)}\,:\,(u_{(i,j)}) \,\rightarrow\,\left(\Phi(n+i,m+j,u_{(i,j)};\varepsilon)\right)\,.
\label{eq:gract} \end{equation}

The infinitesimal generator of the group action of $G$ on the domain of the dependent variable is given by the vector field
$${\mathbf{x}}\,=\,R(n,m,u_{(0,0)})\,\partial_{u_{(0,0)}}\,,$$
where {\em the symmetry characteristic} $R(n,m,u_{(0,0)})$ is defined by
$$R(n,m,u_{(0,0)})\,=\, \left . \frac{{\rm{d}} \phantom{\varepsilon}}{{\rm{d}} \varepsilon} \Phi(n,m,u_{(0,0)};\varepsilon) \right| _{\varepsilon = 0}\,.$$
The group action is reconstructed by exponentiating that of the vector field $\mathbf{x}$
$$\Phi(n,m,u_{(0,0)};\varepsilon)\,=\,\exp(\varepsilon {\mathbf{x}}) \,u_{(0,0)}\,.$$
The infinitesimal generator of the action of $G^{(k)}$ on ${\rm{J}}^{(k)}$ is the associated $k^{\rm{th}}$ order forward prolonged vector field
$${\mathbf{x}}^{(k)}\,=\,\sum_{i=0}^{k}\sum_{j=0}^{k-i} \left({\mathcal{S}}_n^{(i)}\circ {\mathcal{S}}_m^{(j)} R\right)(n,m,u_{(0,0)}) \,\partial_{u_{(i,j)}}\,.$$

The transformation group $G$ is a Lie-point symmetry of the lattice equation (\ref{eq:eq})
if it transforms any solution of (\ref{eq:eq}) to another solution of the same equation. 
Equivalently, $G$ is a symmetry of equation (\ref{eq:eq}),
if the latter is not affected by the transformation (\ref{eq:gract}). 
The infinitesimal criterion for a connected group of transformation $G$ to be
a symmetry  of equation (\ref{eq:eq}) is
\begin{equation}
{\mathbf{x}}^{(k)} \left({\mathcal{E}}\left(n,m,u_{(0,0)},u_{(1,0)},u_{(0,1)},\ldots\right) \right)\,=\,0\,.
\label{eq:infcr} \end{equation}
This should hold for all solutions of equation (\ref{eq:eq}) and, thus, the latter and its consequences should be taken into account. 
Equation (\ref{eq:infcr}) delivers the most general infinitesimal Lie point symmetry of equation (\ref{eq:eq}). The resulting set of
infinitesimal generators forms a Lie algebra $\frak{g}$ from which the corresponding symmetry group $G$ can be constructed by exponentiation.

A {\em lattice invariant} under the action of $G$ is a function $I:{\rm J}^{(k,-k)}\rightarrow \mathds{C}$ 
which satisfies $I(g^{(k)} \cdot (u_{(\pm i,\pm j)})) = I(u_{(\pm i,\pm j)})$ for all $g \in G$ and all 
$(u_{(\pm i,\pm j)}) \in {\rm J}^{(k,-k)}$.
For connected groups of transformations, a necessary and sufficient condition for a function 
$I$ to be invariant under the action of $G$ is the annihilation 
of $I$ by all prolonged infinitesimal generators, i.e.
\begin{equation}
\mathbf{{x}}^{(k,-k)}(I) = 0 \,, \label{eq:pde}
\end{equation}
for all $\mathbf{{x}} \in \mathfrak{g}$.

By relaxing the geometric assumption that the symmetry characteristic $R$ depends on $n$, $m$ and $u_{(0,0)}$, only,
and allowing $R$ to be a function defined on $\mathds{Z}^2\times{\rm J}^{(k,-k)}$ for some finite but unspecified 
$k\in \mathds{N}$, $k\ge 1$, we arrive naturally at the notion of the generalized Lie symmetry. Symmetry generators of this type cannot be associated to transformation groups acting geometrically on the domain of the dependent variable. Lowest order $(k=1)$ generalized symmetries are given by the following vector field
$${\mathbf{v}}\,=\,R(n,m,u_{(0,0)},u_{(1,0)},u_{(0,1)},u_{(-1,0)},u_{(0,-1)}) \partial_{u_{(0,0)}}\,.$$

\section{A class of two-dimensional lattice equations}\label{sec3}

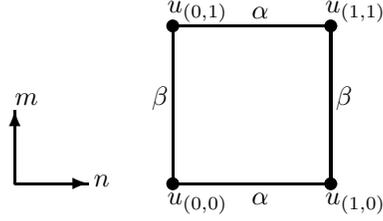
\begin{figure}[h]
\begin{center}
\setlength{\unitlength}{.7cm}%
\begingroup\makeatletter\ifx\SetFigFont\undefined%
\gdef\SetFigFont#1#2#3#4#5{%
  \reset@font\fontsize{#1}{#2pt}%
  \fontfamily{#3}\fontseries{#4}\fontshape{#5}%
  \selectfont}%
\fi\endgroup%
\begin{picture}(3.5,3.7)(0,0) 
\thicklines 
\put(0,0){\line(0,0){3}}
\put(0,3){\line(1,0){3}}
\put(3,0){\line(0,0){3}}
\put(0,0){\line(1,0){3}}
\put(-3,0){\vector(0,0){1.4}}
\put(-3,0){\vector(1,0){1.4}}
\put(0,0){\circle*{.2}}
\put(0,3){\circle*{.2}}
\put(3,0){\circle*{.2}}
\put(3,3){\circle*{.2}}
\put(-1.5,-.05){\makebox(0,0)[lb]{\smash{\SetFigFont{10}{12}{\rmdefault}{\mddefault}{\updefault}$n$}}}
\put(-3,1.5){\makebox(0,0)[lb]{\smash{\SetFigFont{10}{12}{\rmdefault}{\mddefault}{\updefault}$m$}}}
\put(-.1,-.40){\makebox(0,0)[lb]{\smash{\SetFigFont{10}{12}{\rmdefault}{\mddefault}{\updefault}$u_{(0,0)}$}}}
\put(2.9,-0.40){\makebox(0,0)[lb]{\smash{\SetFigFont{10}{12}{\rmdefault}{\mddefault}{\updefault}$u_{(1,0)}$}}}
\put(-.1,3.25){\makebox(0,0)[lb]{\smash{\SetFigFont{10}{12}{\rmdefault}{\mddefault}{\updefault}$u_{(0,1)}$}}}
\put(2.9,3.25){\makebox(0,0)[lb]{\smash{\SetFigFont{10}{12}{\rmdefault}{\mddefault}{\updefault}$u_{(1,1)}$}}}
\put(1.5,-.4){\makebox(0,0)[lb]{\smash{\SetFigFont{10}{12}{\rmdefault}{\mddefault}{\updefault}$\alpha$}}}
\put(-.4,1.5){\makebox(0,0)[lb]{\smash{\SetFigFont{10}{12}{\rmdefault}{\mddefault}{\updefault}$\beta$}}}
\put(1.5,3.15){\makebox(0,0)[lb]{\smash{\SetFigFont{10}{12}{\rmdefault}{\mddefault}{\updefault}$\alpha$}}}
\put(3.1,1.5){\makebox(0,0)[lb]{\smash{\SetFigFont{10}{12}{\rmdefault}{\mddefault}{\updefault}$\beta$}}}
\end{picture}
\vspace{.3cm}
\caption{{\em{An elementary quadrilateral}}}
 \label{fig:quad} \end{center}
\end{figure}

In this section, we present a class of two-dimensional lattice equations, which involve the values of a function $u$ at the vertices of an elementary quadrilateral as shown in Figure \ref{fig:quad}. 
Specifically, we consider the two-dimensional lattice equations of the form
\begin{equation}
Q(u_{(0,0)},u_{(1,0)},u_{(0,1)},u_{(1,1)};\alpha,\beta) \,=\, 0\,, \label{eq:genform}
\end{equation}
 where the function $Q$ 
\begin{itemize}
\item does not depend explicitly on the discrete variables $n$, $m$,
\item depends explicitly on the values of the unknown function $u$ at the vertices of an elementary quadrilateral, 
i.e. $\partial_{u_{(i,j)}} Q(u_{(0,0)},u_{(1,0)},u_{(0,1)},u_{(1,1)};\alpha,\beta) \ne 0$, where $i$, $j$ = 0, 1, and may depend on the parameters $\alpha$, $\beta$ of the lattice,
\item is linear in each argument (affine linear): $\partial_{u_{(i,j)}}^2 Q(u_{(0,0)},u_{(1,0)},u_{(0,1)},u_{(1,1)};\alpha,\beta) = 0$, where $i$, $j$ = 0, 1,
\item and possesses the symmetries of the square (${\mathrm{D}}_4$-symmetry):
\begin{eqnarray*}
Q(u_{(0,0)},u_{(1,0)},u_{(0,1)},u_{(1,1)};\alpha,\beta) &=&  \epsilon Q(u_{(0,0)},u_{(0,1)},u_{(1,0)},u_{(1,1)};\beta,\alpha) \\
&=&  \sigma Q(u_{(1,0)},u_{(0,0)},u_{(1,1)},u_{(0,1)};\alpha,\beta) \,,
\end{eqnarray*}
where $\epsilon = \pm 1$ and $\sigma = \pm 1$.
\end{itemize}

The symmetry analysis of the above class of equations is significantly simplified by the use of certain polynomials arising from the function $Q$. In the rest of this section, we define these polynomials and derive some of their properties, in order to make the symmetry analysis of the following sections more concise.

To begin with, we note that the linearity of the function $Q$ implies that the functions
\begin{subequations}
\begin{eqnarray}
h(u_{(0,0)},u_{(1,0)};\alpha,\beta) & = & Q\, Q_{,u_{(0,1)} u_{(1,1)}} - Q_{,u_{(0,1)}} Q_{,u_{(1,1)}}\,, \label{defh}\\
h_1(u_{(0,0)},u_{(0,1)};\alpha,\beta)& = & Q\, Q_{,u_{(1,0)} u_{(1,1)}} - Q_{,u_{(1,0)}} Q_{,u_{(1,1)}}\,, \label{defH} \\
h_2(u_{(0,1)},u_{(1,1)};\alpha,\beta) & = & Q\, Q_{,u_{(0,0)} u_{(1,0)}} - Q_{,u_{(0,0)}} Q_{,u_{(1,0)}}\,, \label{defhh}\\
h_3(u_{(1,0)},u_{(1,1)};\alpha,\beta) & = & Q\, Q_{,u_{(0,0)} u_{(0,1)}} - Q_{,u_{(0,0)}} Q_{,u_{(0,1)}}\,, \label{defHt}
\end{eqnarray} \label{defhH}
\end{subequations}
are biquadratic polynomials in their two first indicated arguments, and the same holds for the functions
\begin{subequations}
\begin{eqnarray}
G(u_{(0,0)},u_{(1,1)};\alpha,\beta)  &=& Q Q_{,u_{(1,0)} u_{(0,1)}} - Q_{,u_{(1,0)}} Q_{,u_{(0,1)}} \,, \label{defG}\\
G_1(u_{(1,0)},u_{(0,1)};\alpha,\beta)  &=&  Q Q_{,u_{(0,0)} u_{(1,1)}} - Q_{,u_{(0,0)}} Q_{,u_{(1,1)}}\,. \label{defG1}
\end{eqnarray}
\label{defGG1} 
\end{subequations}
In fact, the linearity of the function $Q$ and the above definitions lead immediately to the properties expressed by the following two lemmas.

\begin{lm}
Let the function $Q$ be affine linear. The polynomials defined by (\ref{defhH}), (\ref{defGG1}) are constants if and only if the function $Q$ is linear, i.e.
\begin{equation}
Q\,=\,f_1(\alpha,\beta) u_{(0,0)} + f_2(\alpha,\beta) u_{(1,0)} + f_3(\alpha,\beta) u_{(0,1)} + f_4(\alpha,\beta) u_{(1,1)} + f_5(\alpha,\beta)\,. \label{linear}
\end{equation}
\end{lm}
{\bf{Proof :}} If the function $Q$ is of the form (\ref{linear}, then definitions (\ref{defhH}), (\ref{defGG1}) imply that these polynomials are constants.
Conversely, assuming that these polynomials are constants, we solve (\ref{defhH}), (\ref{defGG1}) for the second order derivatives of $Q$ and take the compatibility conditions among the resulting equations. This leads to eight first-order partial differential equations for the function $Q$. This overdetermined system of partial differential equations imply that $Q$ is necessarily of the form (\ref{linear}). \hfill $\Box$

\begin{lm} If the function $Q$ is affine linear, then the relations
\begin{eqnarray} h(u_{(0,0)},u_{(1,0)};\alpha,\beta)\,h_2(u_{(1,0)},u_{(1,1)};\alpha,\beta)\,&=&\,h_1(u_{(0,0)},u_{(0,1)};\alpha,\beta)\,h_3(u_{(1,0)},u_{(1,1)};\alpha,
\beta)\nonumber \\
&=& \,G(u_{(0,0)},u_{(1,1)};\alpha,\beta)\,G_1(u_{(1,0)},u_{(0,1)};\alpha,\beta) \label{relgenhHG}
\end{eqnarray}
hold, in view of the equation $Q(u_{(0,0)},u_{(1,0)},u_{(0,1)},u_{(1,1)};\alpha,\beta)=0$. 
\end{lm}
{\bf{Proof :}} It follows from the definitions of the functions involved, the affine linearity of the function 
$Q$ and by taking into account the equation $Q=0$.\hfill $\Box$

The ${\mathrm{D}}_4$-symmetry of $Q$ implies that the polynomials defined by (\ref{defhH}) are symmetric in their first  two arguments and related as follows
\begin{equation}
 h(u_{(0,0)},u_{(1,0)};\alpha,\beta) =h_1(u_{(0,0)},u_{(1,0)};\beta,\alpha) = h_2(u_{(0,0)},u_{(1,0)};\alpha,\beta) =h_3(u_{(0,0)},u_{(1,0)};\beta,\alpha).
\label{eq:equivh}
\end{equation}
Moreover, the functions $G$ and $G_1$, defined by (\ref{defGG1}), are symmetric in their first pair of arguments and in $(\alpha,\beta)$, and they have the same form, i.e.
\begin{equation}
G_1(u_{(1,0)},u_{(0,1)};\alpha,\beta) \,=\,G(u_{(1,0)},u_{(0,1)};\alpha,\beta)\,.\label{eq:equivG}
\end{equation}
In this case, Equation (\ref{relgenhHG}) simplifies to
\begin{eqnarray}
h(u_{(0,0)},u_{(1,0)})\,h(u_{(0,1)},u_{(1,1)})\,&=& \,h(u_{(0,0)},u_{(0,1)})\,h(u_{(1,0)},u_{(1,1)})\nonumber\\
&=&\,G(u_{(0,0)},u_{(1,1)})\,G(u_{(1,0)},u_{(0,1)})\,. \label{relhHG}
\end{eqnarray}

\begin{rem} In the following, we omit the dependence of the polynomials $h$, $G$ on the lattice parameters. It should be noted that when the polynomial $h$ is evaluated at two neighboring points in the horizontal direction of the lattice, then the parameter dependence is $(\alpha,\beta)$ and the order is reversed when $h$ involves two points in the vertical direction.\end{rem}

\begin{rem} 
If the function $Q$ is affine linear and possesses the ${\mathrm{D}}_4$-symmetry, then,
following the proof of Lemma 2, one can prove the validity of the following relations
\begin{subequations}
\begin{eqnarray}
Q_{,u_{(0,0)}}^2 + \frac{h(u_{(1,0)},u_{(1,1)}) G(u_{(1,0)},u_{(0,1)})}{h(u_{(0,0)},u_{(1,0)})} &=& 0 \,,\label{rel1} \\
Q_{,u_{(1,1)}}^2 + \frac{h(u_{(0,0)},u_{(0,1)}) G(u_{(1,0)},u_{(0,1)})}{h(u_{(0,1)},u_{(1,1)})} &=& 0 \,, \label{rel2} \\
\frac{Q_{,u_{(1,0)}}}{Q_{,u_{(1,1)}}} - \frac{h(u_{(0,1)},u_{(1,1)})}{G(u_{(1,0)},u_{(0,1)})} &=& 0 \,, \label{rel3} \\
\frac{Q_{,u_{(0,1)}}}{Q_{,u_{(0,0)}}} - \frac{h(u_{(0,0)},u_{(1,0)})}{G(u_{(1,0)},u_{(0,1)})} &=&  0 \,. \label{rel4}
\end{eqnarray}
\label{relall}
\end{subequations}
The latter are quite useful in the symmetry analysis of the equations under consideration.
\end{rem}

Relations (\ref{relhHG}) hold, in general, in view of the equation $Q=0$. However, in certain cases, these relations hold identically, i.e. without taking into account the equation $Q=0$. In such cases, the corresponding equations can be linearized using an appropriate transformation.

\begin{thm}
Let the function $Q$ be affine linear and possess the $\mathrm{D}_4$-symmetry. If the relation
\begin{equation} h(u_{(0,0)},u_{(1,0)}) h(u_{(0,1)},u_{(1,1)}) - G(u_{(0,0)},u_{(1,1)}) G(u_{(1,0)},u_{(0,1)}) \,=\,0 \label{rellin} \end{equation}
holds identically, i.e. without taking into account the equation $Q=0$, then the polynomials $h$, $G$ are factorized as
\begin{eqnarray*}
h(u_{(0,0)},u_{(1,0)};\alpha,\beta) \, &= &\, p(u_{(0,0)};\alpha,\beta)p(u_{(1,0)};\alpha,\beta)\,,\\
G(u_{(1,0)},u_{(0,1)};\alpha,\beta)\,& =& \,\pm p(u_{(1,0)};\alpha,\beta)p(u_{(0,1)};\alpha,\beta)\,,
\end{eqnarray*}
and the equation $Q=0$ is transformed to a linear equation under the transformation
$$u \,\longrightarrow \, {\tilde{u}} \,=\, {\cal{T}}(u)\, := \, \int\frac{1}{p(u;\alpha,\beta)} {\mbox{d}}u\,.$$
\end{thm}
{\bf{Proof :}} First, we write equation (\ref{rellin}) in the form
$$\frac{h(u_{(0,0)},u_{(1,0)})}{G(u_{(1,0)},u_{(0,1)})} = \frac{G(u_{(0,0)},u_{(1,1)})}{h(u_{(0,1)},u_{(1,1)})}\,.$$
Differentiating this relation w.r.t. $u_{(1,0)}$ (equivalently w.r.t. $u_{(1,1)}$), we find that the polynomials $h$, $G$ must have the form
\begin{eqnarray*}
h(u_{(0,0)},u_{(1,0)};\alpha,\beta) \, &= &\, p(u_{(0,0)};\alpha,\beta)p(u_{(1,0)};\alpha,\beta)\,,\\
G(u_{(1,0)},u_{(0,1)};\alpha,\beta)\,& =& \,\pm p(u_{(1,0)};\alpha,\beta)p(u_{(0,1)};\alpha,\beta)\,,
\end{eqnarray*}
where $p$ is a quadratic polynomial in its first argument and symmetric in $(\alpha,\beta)$. 

Now let
$$\tilde{u}\,=\,{\cal{T}}(u) \,:=\,\int\frac{1}{p(u;\alpha,\beta)}\,{\rm{d}}u\,,$$
and
$$F(\tilde{u}_{(0,0)},\tilde{u}_{(1,0)},\tilde{u}_{(0,1)},\tilde{u}_{(1,1)}) \,=\,Q \cdot \left(p(u_{(0,0)};\alpha,\beta) p(u_{(1,0)};\alpha,\beta) p(u_{(0,1)};\alpha,\beta) p(u_{(1,1)};\alpha,\beta)\right)^{-1/2}\,.$$
Using the previous result, it immediately follows that, under the transformation $(u,Q) \longrightarrow (\tilde{u},F)$, the partial differential equations  (\ref{defhH}), (\ref{defGG1}) simplify to the following system
\begin{eqnarray*}
F F_{,\tilde{u}_{(0,1)} \tilde{u}_{(1,1)}} - F_{,\tilde{u}_{(0,1)}} F_{,\tilde{u}_{(1,1)}} \,&=&\, 1\,,\\
F F_{,\tilde{u}_{(1,0)} \tilde{u}_{(1,1)}} - F_{,\tilde{u}_{(1,0)}} F_{,\tilde{u}_{(1,1)}} \,&=&\, 1\,, \\
F F_{,\tilde{u}_{(0,0)} \tilde{u}_{(1,0)}} - F_{,\tilde{u}_{(0,0)}} F_{,\tilde{u}_{(1,0)}} \,&=&\, 1\,, \\
F F_{,\tilde{u}_{(0,0)} \tilde{u}_{(0,1)}} - F_{,\tilde{u}_{(0,0)}} F_{,\tilde{u}_{(0,1)}} \,&=&\, 1\,, \\
F F_{,\tilde{u}_{(1,0)} \tilde{u}_{(0,1)}} - F_{,\tilde{u}_{(1,0)}} F_{,\tilde{u}_{(0,1)}} \,&=&\, \pm 1\,, \\
F F_{,\tilde{u}_{(0,0)} \tilde{u}_{(1,1)}} - F_{,\tilde{u}_{(0,0)}} F_{,\tilde{u}_{(1,1)}} \,&=&\, \pm 1\,.
\end{eqnarray*}
From this it follows that the equation $F(\tilde{u}_{(0,0)},\tilde{u}_{(1,0)},\tilde{u}_{(0,1)},\tilde{u}_{(1,1)})=0$ must be linear. \hfill $\Box$

\begin{ex} {\rm{ Consider the equation
$$u_{(0,0)} \left( u_{(1,0)} u_{(0,1)}+ u_{(1,0)} u_{(1,1)} + u_{(0,1)} u_{(1,1)}\right) + u_{(1,0)} u_{(0,1)} u_{(1,1)} + u_{(0,0)} + u_{(1,0)} + 
u_{(0,1)} + u_{(1,1)}=0,$$
listed in \cite{Hiet}. One easily finds that, in this case,
$$h(u_{(0,0)},u_{(1,0)})\,=\,-(u_{(0,0)}^2-1) (u_{(1,0)}^2-1)\,,\,\,\,G(u_{(1,0)},u_{(0,1)})\,=\,-(u_{(1,0)}^2-1) (u_{(0,1)}^2-1)\,.$$
Preforming the transformation
$$u \longrightarrow \tilde{u}\,=\,\int\frac{{\mbox{d}}u}{u^2-1}\,=\,\frac{1}{2} 
\log\left(\frac{u-1}{u+1}\right)\,,$$
or, equivalently,
$$u\,=\,\frac{1+{\mbox{e}}^{2 \tilde{u}}}{1-{\mbox{e}}^{2 \tilde{u}}}\,,$$
the above equation linearizes to
$$\tilde{u}_{(0,0)} + \tilde{u}_{(1,0)} + \tilde{u}_{(0,1)} + \tilde{u}_{(1,1)}\,=\,0\,.$$}}
\hfill $\Box$
\end{ex}

On the other hand, if the relations (\ref{relhHG} do not hold identically, then the equation $Q=0$ can be written as a conservation law. Specifically,

\begin{thm}
Let the function $Q$ be affine linear and possess the $\mathrm{D}_4$-symmetry. If the relation
\begin{equation}
h(u_{(0,0)},u_{(1,0)})\,h(u_{(0,1)},u_{(1,1)})\,=\,h(u_{(0,0)},u_{(0,1)})\,h(u_{(1,0)},u_{(1,1)})
\label{conscon}
\end{equation}
does not hold identically, then the equation $Q=0$ can be written in the form of a non-trivial conservation law 
\begin{equation}
\left({\mathcal{S}}_m - id \right)\,F_1(n,m,u_{(0,0)},u_{(1,0)})\,=\,\left({\mathcal{S}}_n - id \right)\,F_2(n,m,u_{(0,0)},u_{(0,1)})\,, \label{conslaw}
\end{equation}
where
\begin{subequations}
\begin{eqnarray}
F_1(n,m,u_{(0,0)},u_{(1,0)}) & = & (-1)^{n+m}\,\ln h(u_{(0,0)},u_{(1,0)})\,,\\
F_2(n,m,u_{(0,0)},u_{(0,1)}) & = & (-1)^{n+m}\,\ln h(u_{(0,0)},u_{(0,1)})\,.
\end{eqnarray} \label{consF}
\end{subequations}
\end{thm}
{\bf{Proof :}} It follows by combining (\ref{conslaw}) with (\ref{consF}) and taking into account relation (\ref{conscon}). \hfill $\Box$

\section{Lie point symmetries} \label{lie}

In this section we study the Lie point symmetries of the two-dimensional lattice equations under consideration. Let 
$${\mathbf{x}}\,=\,\phi(n,m,u_{(0,0)})\,\partial_{u_{(0,0)}}$$  
be the generator of a point symmetry transformation of the equation $$Q(u_{(0,0)},u_{(1,0)},u_{(0,1)},u_{(1,1)};\alpha,\beta)\,=\,0\,.$$
The infinitesimal symmetry criterion
$${\mathbf{x}}^{(2)}\left. \left( Q(u_{(0,0)},u_{(1,0)},u_{(0,1)},u_{(1,1)};\alpha,\beta)\right)\right|_{Q=0}\,=\,0\,,$$ 
where
$${\mathbf{x}}^{(2)} \, = \, \sum_{i=0}^{1} \sum_{j=0}^{1} \phi(n+i,m+j,u_{(i,j)};\alpha,\beta) \,\partial_{u_{(i,j)}}\,,$$
implies that the determining equation
\begin{eqnarray} 
Q_{,u_{(0,0)}} \phi(n,m,u_{(0,0)};\alpha,\beta)\, +\, Q_{,u_{(1,0)}} \phi(n+1,m,u_{(1,0)};\alpha,\beta)\,& & \nonumber\\
+\,Q_{,u_{(0,1)}} \phi(n,m+1,u_{(0,1)};\alpha,\beta)\, +\, Q_{,u_{(1,1)}} \phi(n+1,m+1,u_{(1,1)};\alpha,\beta) \,&=&\,0\,, \label{eq:DEqpoint1} 
\end{eqnarray}
should hold on every solution of the equation $Q=0$.

Since the function $Q$ is linear in $u_{(1,1)}$, the equation $Q=0$ can be uniquely solved for $u_{(1,1)}$ in terms of $u_{(0,0)}$, $u_{(1,0)}$ and $u_{(0,1)}$. Using the relations
\begin{equation}
\frac{Q_{,u_{(1,0)}}}{Q_{,u_{(0,0)}}}\,=\,\frac{h(u_{(0,0)},u_{(0,1)})}{G(u_{(1,0)},u_{(0,1)})},\,
\frac{Q_{,u_{(0,1)}}}{Q_{,u_{(0,0)}}}\,=\, \frac{h(u_{(0,0)},u_{(1,0)})}{G(u_{(1,0)},u_{(0,1)})},\,
\frac{Q_{,u_{(1,1)}}}{Q_{,u_{(0,0)}}}\,=\,-\,\frac{Q_{,u_{(1,1)}}^2}{G(u_{(1,0)},u_{(0,1)})}\,,
\end{equation}
we eliminate $u_{(1,1)}$ from equation (\ref{eq:DEqpoint1}) and arrive at
\begin{eqnarray} 
\phi(n,m,u_{(0,0)};\alpha,\beta) \,+\, \frac{h(u_{(0,0)},u_{(0,1)})}{G(u_{(1,0)},u_{(0,1)})} \phi(n+1,m,u_{(1,0)};\alpha,\beta) \nonumber \\
 + \frac{h(u_{(0,0)},u_{(1,0)})}{G(u_{(1,0)},u_{(0,1)})}\phi(n,m+1,u_{(0,1)};\alpha,\beta) \,=\, \frac{Q_{,u_{(1,1)}}^2}{G(u_{(1,0)},u_{(0,1)})} \phi(n+1,m+1,u_{(1,1)};\alpha,\beta) \,. \label{eq:DEqpoint3} \end{eqnarray}
Since $h$ is, in general, a quadratic polynomial of $u_{(0,0)}$, differentiating the determining equation (\ref{eq:DEqpoint3}) three times w.r.t. $u_{(0,0)}$ and once w.r.t. $u_{(1,0)}$, we arrive at
\begin{equation} {\mbox{D}}_{u_{(0,0)}}^3 \,{\mbox{D}}_{u_{(1,0)}}\,\left( \frac{Q_{,u_{(1,1)}}^2}{G(u_{(1,0)},u_{(0,1)})} \,\phi(n+1,m+1,u_{(1,1)};\alpha,\beta) \right) \,=\,0\,, \label{eq:DEqpoint4} \end{equation}
where ${\mbox{D}}_j$ denotes the total derivative operator, i.e.
$${\mbox{D}}_j \,= \,\partial_j \,+\, \left( \partial_j u_{(1,1)} \right)\, \partial_{u_{(1,1)}}\,,\,\,\,\,\mbox{where}\,\,\,j\,=\,u_{(0,0)},\,u_{(1,0)},\,u_{(0,1)}\,.$$

Writing equation (\ref{eq:DEqpoint4}) explicitly, one arrives at
\begin{equation} \partial_{u_{(1,1)}}\,\left(h(u_{(0,1)},u_{(1,1)})^2 \,\partial_{u_{(1,1)}}^3\,\phi(n+1,m+1,u_{(1,1)};\alpha,\beta)\right)\,=\,0\,. \label{eq:DEqpoint5} \end{equation}
The last equation splits into the following system
\begin{eqnarray}
h(0,u_{(1,1)}) \partial_{u_{(1,1)}}^4 \phi\,+\,2 h_{,u_{(1,1)}}(0,u_{(1,1)}) \partial_{u_{(1,1)}}^3 \phi &=& 0\,,\nonumber\\
h_{,u_{(0,1)}}(0,u_{(1,1)}) \partial_{u_{(1,1)}}^4 \phi\,+\,2 h_{,u_{(0,1)} u_{(1,1)}}(0,u_{(1,1)}) \partial_{u_{(1,1)}}^3 \phi &=& 
0\,,\label{eq:DEqsys1}\\
h_{,u_{(0,1)} u_{(0,1)}}(0,u_{(1,1)}) \partial_{u_{(1,1)}}^4 \phi\,+\,2 h_{,u_{(0,1)} u_{(0,1)} u_{(1,1)}}(0,u_{(1,1)}) \partial_{u_{(1,1)}}^3 \phi 
&=& 0\,, \nonumber
\end{eqnarray}
where we have omitted the arguments of the function $\phi(n+1,m+1,u_{(1,1)};\alpha,\beta)$. If the matrix
\begin{equation} 
{\cal{B}} \,=\,\left(\begin{array}{cc} 
h(0,u_{(1,1)}) & h_{,u_{(1,1)}}(0,u_{(1,1)}) \\
h_{,u_{(0,1)}}(0,u_{(1,1)}) & h_{,u_{(0,1)} u_{(1,1)}}(0,u_{(1,1)}) \\
h_{,u_{(0,1)} u_{(0,1)}}(0,u_{(1,1)}) & h_{,u_{(0,1)} u_{(0,1)} u_{(1,1)}}(0,u_{(1,1)})
\end{array}
 \right) \label{defB}
 \end{equation}
has rank equal to 2, then the system (\ref{eq:DEqsys1}) has the unique solution
$$\partial_z^4 \phi(n+1,m+1,u_{(1,1)};\alpha,\beta)\,=\,\partial_z^3 \phi(n+1,m+1,u_{(1,1)};\alpha,\beta)\,=\,0\,, $$
leading to
$$\phi(n,m,u_{(0,0)};\alpha,\beta) \,=\,A_2(n,m;\alpha,\beta) u_{(0,0)}^2 \,+ \,A_1(n,m;\alpha,\beta) u_{(0,0)} \,+\, A_0(n,m;\alpha,\beta) \,.$$

In the degenerate case, where $\mbox{rank} {\cal{B}}\,=\,1$, the function $h(u_{(0,1)},u_{(1,1)})$ separates variables, i.e.
\begin{equation} h(u_{(0,1)},u_{(1,1)})\,=\,h_0(u_{(0,1)}) h_0(u_{(1,1)})\,,\label{eq:DEqpoint6} \end{equation}
where the function $h_0$ is, in general, a quadratic polynomial of its argument. In this case, the system (\ref{eq:DEqsys1}) reduces to one single equation, namely
$$\partial_{u_{(1,1)}} \left(h_0(u_{(1,1)})^2 \partial_{u_{(1,1)}}^3 \phi(n+1,m+1,u_{(1,1)};\alpha,\beta) \right)\,=\,0\,,$$
which integrated once yields
\begin{equation} \partial_{u_{(1,1)}}^3 \phi(n+1,m+1,u_{(1,1)};\alpha,\beta) \,=\,\frac{A_3(n+1,m+1;\alpha,\beta)}{h_0(u_{(1,1)})^2}\,. \label{eq:DEqpoint7}
\end{equation}
Once the function $Q$ is given, the last equation can be easily solved leading to the general form of the corresponding symmetry characteristic.

The form of the functions $A_i(n,m;\alpha,\beta)$ is obtained in the following way. We first substitute the resulting characteristic of the symmetry generator into equation (\ref{eq:DEqpoint3}) and then use the equation to eliminate $u_{(1,1)}$. Setting equal to zero the coefficients of the different monomials of the remaining variables, we finally arrive at a linear overdetermined system of difference equations for the unknown functions $A_i(n,m;\alpha,\beta)$. The solution of this system delivers the Lie point symmetries of the lattice equation $Q = 0$.

{\begin{rem}
Actually, the above symmetry analysis can be immediately extended to the case where the function $Q$ is affine linear but not ${\mathrm{D}}_4$ symmetric. The procedure followed above leads to, essentially, the same equations. They are the ones obtained from equations (\ref{eq:DEqpoint3}) and (\ref{eq:DEqpoint5}) by making the replacements
\begin{eqnarray*}
&& h(u_{(0,0)},u_{(1,0)}) \rightarrow h(u_{(0,0)},u_{(1,0)};\alpha,\beta)\,,\,\,h(u_{(0,0)},u_{(0,1)}) \rightarrow h_1(u_{(0,0)},u_{(0,1)};\alpha,\beta)\,, \\
&&G(u_{(1,0)},u_{(0,1)})\rightarrow G_1(u_{(1,0)},u_{(0,1)};\alpha,\beta)\,,
\end{eqnarray*}
and
$$h(u_{(0,1)},u_{(1,1)}) \,\rightarrow\,h_2(u_{(0,1)},u_{(1,1)};\alpha,\beta)\,, $$
respectively.
\end{rem}}

\section{Three-point generalized symmetries} \label{3point}

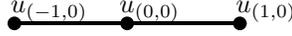
\begin{figure}[h]
\begin{center}
\setlength{\unitlength}{1cm}
\begingroup\makeatletter\ifx\SetFigFont\undefined%
\gdef\SetFigFont#1#2#3#4#5{%
  \reset@font\fontsize{#1}{#2pt}%
  \fontfamily{#3}\fontseries{#4}\fontshape{#5}%
  \selectfont}%
\fi\endgroup%
\begin{picture}(.25,.75)(0,0)
\thicklines
\put(-1,0){\line( 1,0){3}}
\put(-1,0){\circle*{.15}}
\put(.5,0){\circle*{.15}}
\put(2,0){\circle*{.15}}
\put(-1.05,.15){\makebox(0,0)[lb]{\smash{\SetFigFont{10}{12}{\rmdefault}{\mddefault}{\updefault}$u_{(-1,0)}$}}}
\put(.4,.15){\makebox(0,0)[lb]{\smash{\SetFigFont{10}{12}{\rmdefault}{\mddefault}{\updefault}$u_{(0,0)}$}}}
\put(1.95,.15){\makebox(0,0)[lb]{\smash{\SetFigFont{10}{12}{\rmdefault}{\mddefault}{\updefault}$u_{(1,0)}$}}}
\end{picture}
\vspace{.3cm}
\caption{{\em{Three consecutive horizontal lattice points}}} \label{fig:3point}
\end{center}
\end{figure}

In this section, we consider more general symmetries than the point symmetries presented in the preceding section. More precisely, we search for generalized symmetries with characteristics of the form $R(n,m,u_{(0,0)},u_{(1,0)},u_{(-1,0)};\alpha,\beta)$. Once the symmetries of this type are found, one can 
construct similar symmetry characteristics in the other direction of the lattice, simply by interchanging mutually the lattice variables and parameters. This follows from the ${\mathrm{D}}_4$-symmetry of the function $Q(u_{(0,0)},u_{(1,0)},u_{(0,1)},u_{(1,1)};\alpha,\beta)$. Thus, it suffices to restrict our considerations to symmetry characteristics of the form $R(n,m,u_{(0,0)},u_{(1,0)},u_{(-1,0)};\alpha,\beta)$, see Figure \ref{fig:3point}. For brevity, we refer to this type of symmetries as three-point generalized symmetries.

The three-point generalized symmetry analysis for the equations under consideration is summarized in the following
\begin{thm}
Every two-dimensional lattice equation $Q(u_{(0,0)},u_{(1,0)},u_{(0,1)},u_{(1,1)};\alpha,\beta) =0 $, where the function $Q$ is affine linear and possesses the ${\mathrm{D}}_4$-symmetry, admits a three-point generalized symmetry with generator
$${\mbox{\bf{v}}}_n = \left( \frac{h(u_{(0,0)},u_{(1,0)};\alpha,\beta)}{u_{(1,0)}-u_{(-1,0)}}\, -\, \frac{1}{2}
h_{,u_{(1,0)}}(u_{(0,0)},u_{(1,0)};\alpha,\beta) \right) \,\partial_{u_{(0,0)}} \,.$$
Moreover, in the generic case, where the matrix 
\begin{equation}
{\mathcal{G}}\,=\,\left. \left(\begin{array}{ccc}
h(x,y) & G(x,z) & G(x,w) \\
h_{,x}(x,y) & G_{,x}(x,z) & G_{,x}(x,w) \\
h_{,x x}(x,y) & G_{,x x}(x,z) & G_{,x x}(x,w)\end{array}
 \right)\right|_{x\,=\,0} \label{eq:matrixG}
\end{equation}
has rank 3, every three-point generalized symmetry generator necessarily has the form
$${\mbox{\bf{V}}}_{n} =  a(n;\alpha,\beta)\, {\mbox{\bf{v}}}_n  + \frac{1}{2} \phi(n,m,u_{(0,0)};\alpha,\beta) \partial_{u_{(0,0)}}\,,$$
where the functions $a(n;\alpha,\beta)$, $\phi(n,m,u_{(0,0)};\alpha,\beta)$ satisfy equation (\ref{eq:fineq3p}), below.

The ${\mathrm{D}}_4$-symmetry of the function $Q$ implies that the vector fields resulting from the mutual replacements
$$n\,\longleftrightarrow \,m\,,\quad \alpha\,\longleftrightarrow\,\beta\,,\quad u_{(i,0)}\, \longleftrightarrow \,u_{(0,i)}\,,$$
in ${\mbox{\bf{v}}}_n$, ${\mbox{\bf{V}}}_n$, are the generators of the three-point symmetries in the vertical direction.
\label{prop3p} \end{thm} 
\vskip0pt

Since the proof of the Proposition is involved, we give here only a sketch of the proof and leave the details for the Appendix.
\begin{enumerate}
\item We differentiate the determining equation w.r.t. to $u_{(1,-1)}$ and $u_{(1,0)}$ and find that an admissible  symmetry characteristic has the form 
\begin{eqnarray*}
A(n,m;\alpha,\beta) \left(\frac{h(u_{(0,0)},u_{(1,0)};\alpha,\beta)}{u_{(1,0)}-u_{(-1,0)}} - \frac{1}{2} h_{,u_{(1,0)}}(u_{(0,0)},u_{(1,0)};\alpha,\beta)\right)+ \frac{\phi(n,m,u_{(0,0)},u_{(1,0)};\alpha,\beta)}{2},
\end{eqnarray*}
where $A$ and $\phi$ are arbitrary functions of their arguments.
\item Next, we prove that the function $A$ is independent of the discrete variable $m$ and, consequently, the determining equation is independent of $u_{(-1,0)}$ and $u_{(-1,1)}$.
\item Then, we show that the determining equation, apart from the function $\phi$, is also independent of $u_{(2,0)}$.
\item Finally, we eliminate the value $u_{(1,1)}$ from the determining equation, using the equation $Q=0$. At this point, we conclude that every equation, which is affine linear and possesses the symmetries of the square, admits a three-point generalized symmetry with generator
$${\mbox{\bf{v}}}_n = \left( \frac{h(u_{(0,0)},u_{(1,0)};\alpha,\beta)}{u_{(1,0)}-u_{(-1,0)}} - \frac{1}{2}
h_{,u_{(1,0)}}(u_{(0,0)},u_{(1,0)};\alpha,\beta) \right) \partial_{u_{(0,0)}} \,.$$
\item In the generic case, where the matrix $\mathcal{G}$ has rank 3, we prove that the characteristic of the symmetry generator $R(n,m,u_{(0,0)},u_{(1,0)},u_{(-1,0)};\alpha,\beta)$ necessarily has the form 
$$
R \,= \,a(n;\alpha,\beta)\left( \frac{h(u_{(0,0)},u_{(1,0)};\alpha,\beta)}{u_{(1,0)}-u_{(-1,0)}} - \frac{1}{2} h_{,u_{(1,0)}}(u_{(0,0)},u_{(1,0)};\alpha,\beta)\right)+ \frac{\phi(n,m,u_{(0,0)};\alpha,\beta)}{2},
$$
where the functions $a(n;\alpha,\beta)$, $\phi(n,m,u_{(0,0)};\alpha,\beta)$ satisfy the simplified determining equation 
\begin{eqnarray}
\left( a(n;\alpha,\beta) - a(n+1;\alpha,\beta) \right) \,h(u_{(0,0)},u_{(1,0)})^2\,\partial_{u_{(1,0)}} \left( \frac{G(u_{(1,0)},u_{(0,1)})}{h(u_{(0,0)},u_{(1,0)})} \right) && \nonumber \\
&& \nonumber \\
+ G(u_{(1,0)},u_{(0,1)}) \phi(n,m,u_{(0,0)};\alpha,\beta)  + h(u_{(0,0)},u_{(0,1)}) \phi(n+1,m,u_{(1,0)};\alpha,\beta) && \label{eq:fineq3p}\\
&& \nonumber \\
+ h(u_{(0,0)},u_{(1,0)}) \phi(n,m+1,u_{(0,1)};\alpha,\beta) = Q_{,u_{(1,1)}}^2 \phi(n+1,m+1,u_{(1,1)};\alpha,\beta)  \,.&&\nonumber  
\end{eqnarray}
\end{enumerate}

\section{Five-point generalized symmetries} \label{5point}

\begin{figure}[h]
\begin{center}
\setlength{\unitlength}{1cm}%
\begingroup\makeatletter\ifx\SetFigFont\undefined%
\gdef\SetFigFont#1#2#3#4#5{%
  \reset@font\fontsize{#1}{#2pt}%
  \fontfamily{#3}\fontseries{#4}\fontshape{#5}%
  \selectfont}%
\fi\endgroup%
\begin{picture}(3,3.5)(0,0)
\thicklines
\put(1.5,0){\line( 0,0){3}}
\put(0,1.5){\line(1,0){3}}
\put(1.5,1.5){\circle*{.15}}
\put(1.5,0){\circle*{.15}}
\put(1.5,3){\circle*{.15}}
\put(0,1.5){\circle*{.15}}
\put(3,1.5){\circle*{.15}}
\put(0,0){\circle*{.15}}
\put(0,3){\circle*{.15}}
\put(3,3){\circle*{.15}}
\put(3,0){\circle*{.15}}
\put(1.7,1.2){\makebox(0,0)[lb]{\smash{\SetFigFont{10}{12}{\rmdefault}{\mddefault}{\updefault}$u_{(0,0)}$}}}
\put(3.15,1.4){\makebox(0,0)[lb]{\smash{\SetFigFont{10}{12}{\rmdefault}{\mddefault}{\updefault}$u_{(1,0)}$}}}
\put(-1.3,1.4){\makebox(0,0)[lb]{\smash{\SetFigFont{10}{12}{\rmdefault}{\mddefault}{\updefault}$u_{(-1,0)}$}}}
\put(1.4,3.2){\makebox(0,0)[lb]{\smash{\SetFigFont{10}{12}{\rmdefault}{\mddefault}{\updefault}$u_{(0,1)}$}}}
\put(1.4,-.3){\makebox(0,0)[lb]{\smash{\SetFigFont{10}{12}{\rmdefault}{\mddefault}{\updefault}$u_{(0,-1)}$}}}
\end{picture}
\vspace{.3cm}
\caption{{\em{Cross configuration}}} \label{fig:+}
\end{center}
\end{figure}
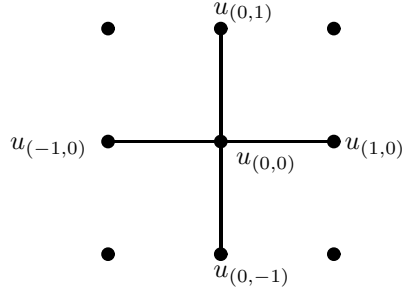

In this section we consider symmetries with characteristic depending on the values of $u$ assigned on five vertices, which form a cross configuration on the lattice as shown in Figure \ref{fig:+}. We refer to this type of symmetries as five-point generalized symmetries. Such symmetry generators exist since any linear combination of the form $c_1 {\mathbf{v}}_n + c_2 {\mathbf{v}}_m$, where ${\mathbf{v}}_n$, ${\mathbf{v}}_m$ are given in the previous section, is also a symmetry generator.

Let
$$
{\mbox{\bf{w}}}^{(2)} =  \sum_{i=0}^{1} \sum_{j=0}^{1}R(n+i,m+j,u_{(i,j)},u_{(i+1,j)},u_{(i,j+1)},u_{(i-1,j)},u_{(i,j-1)};\alpha,\beta)\, \partial_{
u_{(i,j)}}$$
be the prolonged generator of a generalized symmetry of the equation $Q = 0$. Acting on the latter with ${\mbox{\bf{w}}}^{(2)}$ we obtain the determining equation
\begin{eqnarray} 
Q_{,u_{(0,0)}} R(n,m,u_{(0,0)},u_{(1,0)},u_{(0,1)},u_{(-1,0)},u_{(0,-1)};\alpha,\beta)  & & \nonumber \\
+ \,Q_{,u_{(1,0)}} R(n+1,m,u_{(1,0)},u_{(2,0)},u_{(1,1)},u_{(0,0)},u_{(1,-1)};\alpha,\beta) & &  \nonumber  \\
+ \, Q_{,u_{(0,1)}} R(n,m+1,u_{(0,1)},u_{(1,1)},u_{(0,2)},u_{(-1,1)},u_{(0,0)};\alpha,\beta)  & &  \nonumber  \\ 
+ \, Q_{,u_{(1,1)}} R(n+1,m+1,u_{(1,1)},u_{(2,1)},u_{(1,2)},u_{(0,1)},u_{(1,0)};\alpha,\beta) &=& 0\,. \label{eq:15p}
\end{eqnarray}

Since $u_{(2,1)}$ depends on $u_{(2,0)}$ through the equation $Q(u_{(1,0)},u_{(2,0)},u_{(1,1)},u_{(2,1)};\alpha,\beta) = 0$, and $u_{(1,2)}$ depends on $u_{(0,2)}$ through the equation $Q(u_{(0,1)},u_{(1,1)},u_{(0,2)},u_{(1,2)};\alpha,\beta) = 0$, differentiating equation (\ref{eq:15p}) once w.r.t. $u_{(2,0)}$ and then w.r.t. $u_{(0,2)}$, we arrive at
\begin{subequations} \label{eq15p}
\begin{equation}
R_{,u_{(2,1)} u_{(1,2)}}(n+1,m+1, u_{(1,1)},u_{(2,1)},u_{(1,2)},u_{(0,1)},u_{(1,0)};\alpha,\beta) \, \frac{\partial u_{(2,1)}}{\partial u_{(2,0)}} \, \, \frac{\partial u_{(1,2)}}{\partial u_{(0,2)}}\,= \, 0\,.
\end{equation}
On the other hand, $u_{(-1,0)}$ depends on $u_{(-1,1)}$ through the equation $Q(u_{(-1,0)},u_{(0,0)},u_{(-1,1)},u_{(0,1)};\alpha,\beta) = 0$, and $u_{(0,-1)}$ depends on $u_{(1,-1)}$ through the equation $Q(u_{(0,-1)},u_{(1,-1)},u_{(0,0)},u_{(1,0)};\alpha,\beta) = 0$. Thus, the differentiation of equation (\ref{eq:15p}) once w.r.t. $u_{(-1,1)}$ and then w.r.t. $u_{(1,-1)}$ yields 
\begin{equation}
R_{,u_{(-1,0)} u_{(0,-1)}}(n,m, u_{(0,0)},u_{(1,0)},u_{(0,1)},u_{(-1,0)},u_{(0,-1)};\alpha,\beta) \, \frac{\partial u_{(-1,0)}}{\partial u_{(-1,1)}} \, \, \frac{\partial u_{(0,-1)}}{\partial u_{(1,-1)}}\,= \, 0\,.
\end{equation}
Moreover, differentiating equation (\ref{eq:15p}) once w.r.t. $u_{(2,0)}$ (respectively $u_{(0,2)}$) and then w.r.t. $u_{(1,-1)}$ (respectively $u_{(-1,1)}$), we obtain
\begin{eqnarray}
R_{,u_{(2,0)} u_{(1,-1)}}(n+1,m, u_{(1,0)},u_{(2,0)},u_{(1,1)},u_{(0,0)},u_{(1,-1)};\alpha,\beta) &=& 0\,, \\
R_{,u_{(0,2)} u_{(-1,1)}}(n,m+1, u_{(0,1)},u_{(1,1)},u_{(0,2)},u_{(-1,1)},u_{(0,0)};\alpha,\beta) &=& 0\,.
\end{eqnarray}
\end{subequations}

Equations (\ref{eq15p}) imply that
\begin{eqnarray*}
R(n,m,u_{(0,0)},u_{(1,0)},u_{(0,1)},u_{(-1,0)},u_{(0,-1)};\alpha,\beta) &=& R_1(n,m,u_{(0,0)},u_{(1,0)},u_{(-1,0)};\alpha,\beta) \,+ \\
& &R_2(n,m,u_{(0,0)},u_{(0,1)},u_{(0,-1)};\alpha,\beta) \,.
\end{eqnarray*}
Substituting the above relation into the determining equation (\ref{eq:15p}), and following the steps of the proof in Section \ref{3point}, we find that the function $R_1$ has the form
\begin{eqnarray*}
R_1(n,m,u_{(0,0)},u_{(1,0)},u_{(-1,0)};\alpha,\beta) & = & A(n,m;\alpha,\beta) \, \frac{h(u_{(0,0)},u_{(1,0)};\alpha,\beta)}{u_{(1,0)}-u_{(-1,0)}}  
 + \, \phi_1(n,m,u_{(0,0)},u_{(1,0)};\alpha,\beta)\,,
\end{eqnarray*}
where $A$, $\phi_1$ are arbitrary functions of their arguments. In a similar manner, one finds that
\begin{eqnarray*}
R_2(n,m,u_{(0,0)},u_{(0,1)},u_{(0,-1)};\alpha,\beta) & = & B(n,m;\alpha,\beta) \, \frac{h(u_{(0,0)},u_{(0,1)};\beta,\alpha)}{u_{(0,1)}-u_{(0,-1)}} + \, \phi_2(n,m,u_{(0,0)},u_{(0,1)};\alpha,\beta)\,,
\end{eqnarray*}
where $B$, $\phi_2$ are arbitrary functions of their arguments. 

Setting
\begin{eqnarray*}
 \phi_1 \,=\, -\,\frac{1}{2}\, A(n,m;\alpha,\beta)\, h_{,u_{(1,0)}}(u_{(0,0)},u_{(1,0)};\alpha,\beta)\, +\, \frac{1}{2} \,\psi_1(n,m,u_{(0,0)},u_{(1,0)};\alpha,\beta)\,, \\
 \phi_2 \,=\, -\,\frac{1}{2} \,B(n,m;\alpha,\beta)\, h_{,u_{(0,1)}}(u_{(0,0)},u_{(0,1)};\beta,\alpha)\, +\, \frac{1}{2}\, \psi_2(n,m,u_{(0,0)},u_{(0,1)};\alpha,\beta)\,,
\end{eqnarray*}
we substitute the above relations into the determining equation and differentiate the result w.r.t. $u_{(2,0)}$. This leads to the constraint
$$A(n+1,m;\alpha,\beta) - A(n+1,m+1;\alpha,\beta) = 0 \,,$$
which implies that
$$A(n,m;\alpha,\beta) = a(n;\alpha,\beta)\,.$$
Similarly, differentiating the determining equation w.r.t. $u_{(0,2)}$ we arrive at 
$$B(n,m+1;\alpha,\beta) - B(n+1,m+1;\alpha,\beta) = 0\,,$$
which gives
$$B(n,m;\alpha,\beta) = b(m;\alpha,\beta)\,.$$
Again, by differentiation, it is shown that the coefficient of $a(n;\alpha,\beta)$ and $b(m;\alpha,\beta)$ are independent of $u_{(-1,1)}$ and $u_{(1,-1)}$, respectively. Moreover, in the generic case, the rank of the matrix $\mathcal{G}$ implies that the functions $\psi_i$ are independent on their fourth arguments, i.e. $\psi_1(n,m,u_{(0,0)},u_{(1,0)};\alpha,\beta) = \psi_1(n,m,u_{(0,0)};\alpha,\beta)$ and 
$\psi_2(n,m,u_{(0,0)},u_{(0,1)};\alpha,\beta) = \psi_2(n,m,u_{(0,0)};\alpha,\beta)$. In this manner, we arrive at the following form of the determining equation
\begin{eqnarray}
\left( a(n;\alpha,\beta) - a(n+1;\alpha,\beta) \right) \,h(u_{(0,0)},u_{(1,0)})^2\,\partial_{u_{(1,0)}} \left( \frac{G(u_{(1,0)},u_{(0,1)})}{h(u_{(0,0)},u_{(1,0)})} \right) && \nonumber\\
&& \nonumber\\
+ \, \left( b(m;\alpha,\beta) - b(m+1;\alpha,\beta) \right) \,h(u_{(0,0)},u_{(0,1)})^2 \,\partial_{u_{(0,1)}} \left( \frac{G(u_{(1,0)},u_{(0,1)})}{h(u_{(0,0)},u_{(0,1)})} \right) && \nonumber\\
&&  \label{eq:fin5p} \\
+ G(u_{(1,0)},u_{(0,1)}) \psi(n,m,u_{(0,0)};\alpha,\beta)  + h(u_{(0,0)},u_{(0,1)}) \psi(n+1,m,u_{(1,0)};\alpha,\beta) && \nonumber\\
&& \nonumber \\
+ h(u_{(0,0)},u_{(1,0)}) \psi(n,m+1,u_{(0,1)};\alpha,\beta) = Q_{,u_{(1,1)}}^2 \psi(n+1,m+1,u_{(1,1)};\alpha,\beta)  \,,&&
\nonumber \end{eqnarray}
where we have set 
$$\psi_1(n,m,u_{(0,0)};\alpha,\beta)+\psi_2(n,m,u_{(0,0)};\alpha,\beta)\,=\,\psi(n,m,u_{(0,0)};\alpha,\beta)\,,$$
since they appear additively in the characteristic $R$.

The form of the function $\psi$ is obtained in the same way as the one used to obtain the general form of the characteristic of a Lie point symmetry generator. The substitution of the relevant form of the function $\psi$ into the determining equation (\ref{eq:fin5p}) and the usage of equation $Q=0$ to eliminate $u_{(1,1)}$ in the resulting equation yield a polynomial in $u_{(0,0)}$, $u_{(1,0)}$ and $u_{(0,1)}$.
Setting the coefficients of the different monomials equal to zero, we come up with 
an overdetermined linear system of difference equations for the unknown functions $a(n;\alpha,\beta)$, $b(m;\alpha,\beta)$ and the functions $A_i(n,m;\alpha,\beta)$, which occur in the general form of the function $\psi$. The general solution of this system delivers the five-point generalized symmetries, as well as all the three-point generalized and Lie point symmetries. Thus, equation (\ref{eq:fin5p}) is the most general equation for determining the symmetries of a two-dimensional lattice equation, under the specific assumptions. The above analysis summarizes to the following
\begin{thm} 
Consider the equation $Q(u_{(0,0)},u_{(1,0)},u_{(0,1)},u_{(1,1)};\alpha,\beta) =0 $, where function $Q$ is affine linear and possesses the ${\mathrm{D}}_4$ symmetry. In the generic case, where the matrix $\mathcal{G}$ defined by (\ref{eq:matrixG}) has rank 3, the characteristic $R(n,m,u_{(0,0)},u_{(1,0)},u_{(0,1)},u_{(-1,0)},u_{(0,-1)};\alpha,\beta)$ of a five-point symmetry generator has necessarily the form
$$
R\,=\,a(n;\alpha,\beta) P(u_{(0,0)},u_{(1,0)},u_{(-1,0)};\alpha,\beta) + b(m;\alpha,\beta) P(u_{(0,0)},u_{(0,1)},u_{(0,-1)};\beta,\alpha) + \frac{\psi(n,m,u_{(0,0)};\alpha,\beta)}{2},$$
where
$$
P(u,x,y;\alpha,\beta) = \frac{h(u,x;\alpha,\beta)}{x-y} - \frac{1}{2} h_{,x}(u,x;\alpha,\beta) \,,
$$
and the functions $a(n;\alpha,\beta)$, $b(m;\alpha,\beta)$ and $\psi(n,m,u_{(0,0)};\alpha,\beta)$ satisfy equation (\ref{eq:fin5p}). \label{prop5p}
\end{thm}

\section{Extended symmetries on the lattice parameters} \label{param}

It has been observed \cite{Pap2} that certain lattice equations admit compatible systems of differential-difference equations, with the lattice parameters $\alpha$, $\beta$ playing the role of the additional (continuous) independent variables. Such a compatible system arises from the invariance condition of the lattice equation under the action of a generalized symmetry transformation extended to the lattice parameters in a specific way. 

As an example, let us consider the lattice potential KdV equation
$$(u_{(0,0)}-u_{(1,1)})\, (u_{(1,0)}-u_{(0,1)})\, -\,\alpha \,+ \, \beta \, = \,0 \,.$$
Obviously, this equation is not invariant under the scalings 
$$(u_{(0,0)},u_{(1,0)},u_{(0,1)},u_{(1,1)}) \, \rightarrow \,\left({\mbox{e}}^\epsilon u_{(0,0)},{\mbox{e}}^\epsilon u_{(1,0)},{\mbox{e}}^\epsilon u_{(0,1)},{\mbox{e}}^\epsilon u_{(1,1)}\right) \,. $$
However, if it is further assumed that the parameters $\alpha$, $\beta$ change according to
$$(\alpha,\beta) \, \rightarrow \, \left({\mbox{e}}^{2 \epsilon}\,\alpha,{\mbox{e}}^{2 \epsilon} \,\beta \right) \,, $$
then the equation does remain invariant. In other words, the lattice potential KdV  admits the symmetry generated by
$$u_{(0,0)}\,\partial_{u_{(0,0)}} \,+ \,2\,\alpha \partial_\alpha \,+\,2\,\beta \partial_\beta\,.$$

These observations make it clear that it is useful to extend our considerations to symmetry transformations acting on the lattice parameters, as well. In this spirit, the present section is devoted to symmetry transformations of the equations under study, which are generated by vector fields of the form
\begin{equation}
{\mathbf{w}}\,= \,R(n,m,u_{(0,0)},u_{(1,0)},u_{(0,1)},u_{(-1,0)},u_{(0,-1)};\alpha,\beta)\, \partial_{u_{(0,0)}}
+\,\xi(n,m;\alpha,\beta)\,\partial_\alpha \,+\,\zeta(n,m;\alpha,\beta)\,\partial_\beta\,. \label{5p+par}
\end{equation}

Acting with the prolonged symmetry generator on the equation and following the analysis in the preceding sections, we find, in the generic case, that the component in the $u$-direction of $\mathbf{w}$ takes the form
\begin{eqnarray*}
 a(n,m;\alpha,\beta)P(u_{(0,0)},u_{(1,0)},u_{(-1,0)};\alpha,\beta)+b(m;\alpha,\beta) P(u_{(0,0)},u_{(0,1)},u_{(0,-1)};\beta,\alpha) + \frac{1}{2}\psi(n,m,u_{(0,0)};\alpha,\beta),
\end{eqnarray*}
where
$$
P(u,x,y;\alpha,\beta) = \frac{h(u,x;\alpha,\beta)}{x-y} - \frac{1}{2} h_{,x}(u,x;\alpha,\beta) \,,
$$
and the functions $a(n;\alpha,\beta)$, $b(m;\alpha,\beta)$, $\psi(n,m,u_{(0,0)};\alpha,\beta)$, $\xi(n,m;\alpha,\beta)$ and $\zeta(n,m;\alpha,\beta)$ satisfy the determining equation
\begin{eqnarray*}
\left( a(n;\alpha,\beta) - a(n+1;\alpha,\beta) \right) \,h(u_{(0,0)},u_{(1,0)})^2\,\partial_{u_{(1,0)}} \left( \frac{G(u_{(1,0)},u_{(0,1)})}{h(u_{(0,0)},u_{(1,0)})} \right) && \\
&& \\
+ \, \left( b(m;\alpha,\beta) - b(m+1;\alpha,\beta) \right) \,h(u_{(0,0)},u_{(0,1)})^2 \,\partial_{u_{(0,1)}} \left( \frac{G(u_{(1,0)},u_{(0,1)})}{h(u_{(0,0)},u_{(0,1)})} \right) && \\
&&\\
+ G(u_{(1,0)},u_{(0,1)}) \psi(n,m,u_{(0,0)};\alpha,\beta)  + h(u_{(0,0)},u_{(0,1)}) \psi(n+1,m,u_{(1,0)};\alpha,\beta) && \nonumber\\
&& \nonumber \\
+ h(u_{(0,0)},u_{(1,0)}) \psi(n,m+1,u_{(0,1)};\alpha,\beta) - Q_{,u_{(1,1)}}^2 \psi(n+1,m+1,u_{(1,1)};\alpha,\beta) &&\\
&& \\
-\,2 \,Q_{,\alpha}\, Q_{,u_{(1,1)}}\, \xi(n,m;\alpha,\beta)\, -\, 2\, Q_{,\beta}\, Q_{,u_{(1,1)}}\, \zeta(n,m;\alpha,\beta)  = 0 \,.
\nonumber \end{eqnarray*}

Again the form of the function $\psi$ is obtained in the manner described in the previous section, by taking into account that
$${\mbox{D}_{u_{(0,0)}}^3}\, \left(\,Q_{,i} \,Q_{,u_{(1,1)}}\, \right)\,=\,0\,,\,\,\,\,i\,=\,\alpha,\,\beta\,,$$
which follows from the linearity of the function $Q$. The general solution of the above determining equation will give us the Lie point symmetries, the three- and the five-point generalized symmetries.

\section{Symmetries of lattice equations with the consistency property} \label{symabs}

In this section, we apply the results of previous sections to the symmetry analysis to the class of integrable nonlinear equations obtained by Adler, Bobenko and Suris \cite{ABS} recently. Our results are applicable to the ABS equations, because the latter are affine linear and possess the symmetries of the square. They are not linearizable and satisfy the condition ${\mathrm{rank}}\, {\mathcal{G}}=3$, where ${\mathcal{G}}$ is given by (\ref{eq:matrixG}). Moreover, these equations are integrable in the sense that they satisfy the three-dimensional consistency property. The ABS equations are given by entries {\bf{(H1-3)}} and {\bf{(Q1-4)}} of the following list.
\begin{eqnarray} 
{\mbox{\bf{(H1)}}} &\qquad& (u_{(0,0)}-u_{(1,1)})\, (u_{(1,0)}-u_{(0,1)})\, -\,\alpha \,+ \, \beta \, = \,0 \label{H1} \\
&&\nonumber  \\
{\mbox{\bf{(H2)}}} & & (u_{(0,0)}-u_{(1,1)})(u_{(1,0)}-u_{(0,1)}) +(\beta-\alpha) (u_{(0,0)}+u_{(1,0)}+u_{(0,1)}+u_{(1,1)}) \nonumber \\
& & - \alpha^2 + \beta^2 = 0 \label{H2} \\
&& \nonumber \\
{\mbox{\bf{(H3)}}} & & \alpha (u_{(0,0)} u_{(1,0)}+u_{(0,1)} u_{(1,1)}) - \beta (u_{(0,0)} u_{(0,1)}+u_{(1,0)} u_{(1,1)}) + \delta (\alpha^2-\beta^2) = 0 \label{H3} \\
&& \nonumber \\
{\mbox{\bf{(Q1)}}} & &   \alpha (u_{(0,0)}-u_{(0,1)}) (u_{(1,0)}- u_{(1,1)}) - \beta (u_{(0,0)}- u_{(1,0)}) (u_{(0,1)} -u_{(1,1)}) \nonumber \\
& &    + \delta^2 \alpha \beta (\alpha-\beta)= 0 \label{Q1}\\
& & \nonumber \\
{\mbox{\bf{(Q2)}}} & &  \alpha (u_{(0,0)}-u_{(0,1)}) (u_{(1,0)}- u_{(1,1)}) - \beta (u_{(0,0)}- u_{(1,0)}) (u_{(0,1)} -u_{(1,1)}) + \nonumber\\
 &&   \alpha \beta (\alpha-\beta) (u_{(0,0)}+u_{(1,0)}+u_{(0,1)}+u_{(1,1)}) - \alpha \beta (\alpha-\beta) (\alpha^2-\alpha \beta + \beta^2) = 0 \label{Q2}\\
& & \nonumber \\
{\mbox{\bf{(Q3)}}} & &   (\beta^2-\alpha^2) (u_{(0,0)} u_{(1,1)}+u_{(1,0)} u_{(0,1)}) + \beta (\alpha^2-1) (u_{(0,0)} u_{(1,0)}+u_{(0,1)} u_{(1,1)}) \nonumber\\
&&    - \alpha (\beta^2-1) (u_{(0,0)} u_{(0,1)}+u_{(1,0)} u_{(1,1)}) - \frac{\delta^2 (\alpha^2-\beta^2) (\alpha^2-1) (\beta^2-1)}{4 \alpha \beta}=0  \label{Q3} \\
&& \nonumber \\
{\mbox{\bf{(Q4)}}} & &    a_0 u_{(0,0)} u_{(1,0)} u_{(0,1)} u_{(1,1)}  \nonumber \\
&&  + a_1 (u_{(0,0)} u_{(1,0)} u_{(0,1)} + u_{(1,0)} u_{(0,1)} u_{(1,1)} + u_{(0,1)} u_{(1,1)} u_{(0,0)} + u_{(1,1)} u_{(0,0)} u_{(1,0)}) \nonumber \\
& &   +\alpha_2 (u_{(0,0)} u_{(1,1)} + u_{(1,0)} u_{(0,1)}) + \bar{a}_2 (u_{(0,0)} u_{(1,0)}+u_{(0,1)} u_{(1,1)})  \label{Q4} \\
& &  + \tilde{a}_2 (u_{(0,0)} u_{(0,1)}+u_{(1,0)} u_{(1,1)}) + a_3 (u_{(0,0)} + u_{(1,0)} + u_{(0,1)} + u_{(1,1)}) + a_4 = 0\nonumber
\end{eqnarray}
Here $\alpha$, $\beta$ are the lattice parameters and the $a_i$'s in {\bf{(Q4)}} are determined by the relations
\begin{eqnarray*}
&& a_0 = a+b \,,\,\,\,a_1=-a \beta - b \alpha\,,\,\,\,a_2=a \beta^2 + b \alpha^2\,, \\
 \\
&& \bar{a}_2 = \frac{a b (a+b)}{2 (\alpha-\beta)} + a \beta^2 - (2 \alpha^2 - \frac{g_2}{4}) b\,,\,\,
 \tilde{a}_2 = \frac{a b (a+b)}{2 (\beta-\alpha)} + b \alpha^2 - (2 \beta^2 - \frac{g_2}{4}) a\,, \\
 \\
&& a_3 = \frac{g_3}{2}a_0 - \frac{g_2}{4} a_1\,,\,\,\,a_4=\frac{g_2^2}{16}a_0-g_3 a_1\,,
\end{eqnarray*}
with 
$$a^2\, =\, r(\alpha)\,,\quad b^2 \,=\, r(\beta)\,,\quad r(x)\,=\,4 x^3-g_2 x - g_3\,.$$

In fact, the results of Sections \ref{lie}-\ref{5point} allow us to obtain {\sl all} the Lie point symmetries, the three- and five-point generalized symmetries of the ABS equations. The generators of these symmetries are given in the following list. We give the corresponding generators, using the symbol of each equation employed in the previous list.

\begin{itemize} 
\item {\bf{H1}}
 
 {\em{Point symmetries}} : ${\mathbf{x}}_1 = \partial_{u_{(0,0)}},\,{\mathbf{x}}_2 = (-1)^{n-m} \partial_{u_{(0,0)}},\,{\mathbf{x}}_3 =(-1)^{n-m} u_{(0,0)} \partial_{u_{(0,0)}}.$ \\
    
{\em{Three-point generalized symmetries}} :
    \begin{eqnarray*}
    {\mathbf{v}}_1 &=&  \frac{1}{u_{(1,0)}-u_{(-1,0)}} \partial_{u_{(0,0)}} \,,\,\,\,{\mathbf{v}}_{2}\, =\, n\, {\mathbf{v}}_1 \, +\, \frac{u_{(0,0)}}{2 (\alpha-\beta)}\, \partial_{u_{(0,0)}}\,, \\
    & & \\
    {\mathbf{v}}_3 &=&  \frac{1}{u_{(0,1)}-u_{(0,-1)}} \partial_{u_{(0,0)}} \,,\,\,\,{\mathbf{v}}_{4} =\,m\,{\mathbf{v}}_3 - \frac{u_{(0,0)}}{2 (\alpha-\beta)}\, \partial_{u_{(0,0)}}\,.
    \end{eqnarray*}

\item {\bf{H2}} 

{\em{Point symmetries}} : ${\mathbf{x}}_1 = (-1)^{n+m} \partial_{u_{(0,0)}}\,.$\\
    
{\em{Three-point generalized symmetries}} :
    \begin{eqnarray*}
    {\mathbf{v}}_1 &=&  \frac{u_{(1,0)} + 2 u_{(0,0)} + u_{(-1,0)} + 2 \alpha}{u_{(1,0)}-u_{(-1,0)}} \partial_{u_{(0,0)}} \,,\,\,\,{\mathbf{v}}_{2} =  n\,{\mathbf{v}}_1 + \frac{2 u_{(0,0)} + \beta}{2 
(\alpha-\beta)}\, \partial_{u_{(0,0)}}\,, \\
    & & \\
    {\mathbf{v}}_3 &=& \frac{u_{(1,0)} + 2 u_{(0,0)} + u_{(0,-1)} + 2 \beta}{u_{(0,1)}-u_{(0,-1)}} \partial_{u_{(0,0)}} \,,\,\,\,{\mathbf{v}}_{4} = m\,{\mathbf{v}}_3 - \frac{2 u_{(0,0)} + \alpha}{2 
(\alpha-\beta)} \,\partial_{u_{(0,0)}}\,.
    \end{eqnarray*} 

\item {\bf{H3}}
\begin{enumerate}
  \item  $\delta = 0$.
  
       {\em{Point symmetries}} : ${\mathbf{x}}_1 = u_{(0,0)} \partial_{u_{(0,0)}}\,,\,\,\,{\mathbf{x}}_2 =  (-1)^{n+m} u_{(0,0)} \partial_{u_{(0,0)}}\,.$\\
    
    {\em{Three-point generalized symmetries}} :
    \begin{eqnarray*}
    {\mathbf{v}}_1 &=&  \frac{u_{(0,0)} (u_{(1,0)}+ u_{(-1,0)})}{u_{(1,0)}-u_{(-1,0)}} \partial_{u_{(0,0)}} \,,\,\,\,{\mathbf{v}}_{2} = \frac{u_{(0,0)} (u_{(0,1)} + u_{(0,-1)})}{u_{(0,1)}-u_{(0,-1)}} \partial_{u_{(0,0)}}\,.   
    \end{eqnarray*}\\
    
    {\em{Five-point generalized symmetries}} : ${\mathbf{w}} \,=\, n\, {\mathbf{v}}_1 \,+\,m\,{\mathbf{v}}_3 \,.$\\
    
  \item $\delta \ne 0$. 
  
       {\em{Point symmetries}} : $ {\mathbf{x}}_1=  (-1)^{n+m} u_{(0,0)} \partial_{u_{(0,0)}}\,. $\\
    
    {\em{Three-point generalized symmetries}} :
    \begin{eqnarray*}
     {\mathbf{v}}_1 &=&  \frac{u_{(0,0)} (u_{(1,0)}+ u_{(-1,0)}) \,+\, 2\, \alpha\, \delta}{u_{(1,0)}-u_{(-1,0)}} \partial_{u_{(0,0)}} \,,\,\,\,{\mathbf{v}}_{2} = \frac{u_{(0,0)} (u_{(0,1)} + u_{(0,-1)}) \,+\, 2\, \beta\, \delta}{u_{(0,1)}-u_{(0,-1)}} \partial_{u_{(0,0)}}\,. 
    \end{eqnarray*} \\
    
   {\em{Five-point generalized symmetries}} : ${\mathbf{w}} \,=\, n\,{\mathbf{v}}_1 \,+\,m\,{\mathbf{v}}_2\,-\,\frac{u_{(0,0)}}{2}\,\partial_{u_{(0,0)}}\,.$ \\
\end{enumerate}

\item {\bf{Q1}}
  
  \begin{enumerate}
  \item $\delta = 0$ 
  
        {\em{Point symmetries}} : $ {\mathbf{x}}_1=  u_{(0,0)}^2 \partial_{u_{(0,0)}},\,{\mathbf{x}}_2=  u_{(0,0)} \partial_{u_{(0,0)}},\,{\mathbf{x}}_3=  \partial_{u_{(0,0)}}\,.$\\
    
    {\em{Three-point generalized symmetries}} :
    \begin{eqnarray*}
    {\mathbf{v}}_1 &=& \frac{(u_{(1,0)}-u_{(0,0)}) (u_{(0,0)}-u_{(-1,0)})}{u_{(1,0)}-u_{(-1,0)}}\partial_{u_{(0,0)}}\,, \\
    {\mathbf{v}}_2 &=& \frac{(u_{(0,1)}-u_{(0,0)}) (u_{(0,0)}-u_{(0,-1)})}{u_{(0,1)}-u_{(0,-1)}}\partial_{u_{(0,0)}}\,.
    \end{eqnarray*} \\
       
    {\em{Five-point generalized symmetries}} : $ {\mathbf{w}} \,=\, n\,{\mathbf{v}}_1 \,+\,m\,{\mathbf{v}}_2\,.$\\

 \item $\delta \ne 0$ 
 
       {\em{Point symmetries}} : $ {\mathbf{x}}_1= \partial_{u_{(0,0)}}\,. $\\    
       
    {\em{Three-point generalized symmetries}} :
    \begin{eqnarray*}
    {\mathbf{v}}_1 &=& \frac{(u_{(1,0)}-u_{(0,0)}) (u_{(0,0)}-u_{(-1,0)})+ \alpha^2 \delta^2}{u_{(1,0)}-u_{(-1,0)}}\partial_{u_{(0,0)}}\,, \\
    & & \\
    {\mathbf{v}}_2 &=& \frac{(u_{(0,1)}-u_{(0,0)}) (u_{(0,0)}-u_{(0,-1)})+ \beta^2 \delta^2}{u_{(0,1)}-u_{(0,-1)}}\partial_{u_{(0,0)}}\,.
    \end{eqnarray*}\\
       
    {\em{Five-point generalized symmetries}} : $ {\mathbf{w}} \,=\, n\,{\mathbf{v}}_1 \,+\,m\,{\mathbf{v}}_2 \,-\,u_{(0,0)}\,\partial_{u_{(0,0)}}\,.$ \\
\end{enumerate}    
    
     \item {\bf{Q2}}
     
           {\em{Three-point generalized symmetries}}  : 
    \begin{eqnarray*}
     {\mathbf{v}}_1 \,=\, \frac{(u_{(1,0)}-u_{(0,0)}) (u_{(0,0)}-u_{(-1,0)}) + (u_{(1,0)} + 2 u_{(0,0)} + u_{(-1,0)})\alpha^2 - \alpha^4}{u_{(1,0)}-u_{(-1,0)}} \partial_{u_{(0,0)}}\,,\\
    & & \\
    {\mathbf{v}}_2 \,=\, \frac{(u_{(0,1)}-u_{(0,0)}) (u_{(0,0)}-u_{(0,-1)}) + (u_{(0,1)} + 2 u_{(0,0)} + u_{(0,-1)})\beta^2 - \beta^4}{u_{(0,1)}-u_{(0,-1)}} \partial_{u_{(0,0)}}\,.
    \end{eqnarray*} \\
                            
    {\em{Five-point generalized symmetries}} : $ {\mathbf{w}} \,=\, n\,{\mathbf{v}}_1 \,+\,m\,{\mathbf{v}}_2 \,-\,2 u_{(0,0)}\,\partial_{u_{(0,0)}} \,.$ \\

    \item {\bf{Q3}}
    
        {\em{Point symmetries}}  :  If $\delta = 0$, then it admits one point symmetry with generator ${\mbox{\bf{x}}}_1 = u_{(0,0)} \partial_{u_{(0,0)}}$. Otherwise, there are no point symmetries.\\
    
    {\em{Three-point generalized symmetries}} :
    \begin{eqnarray*}
    {\mathbf{v}}_1 \,=\, \frac{2 \alpha (\alpha^2+1) u_{(0,0)} (u_{(1,0)}+u_{(-1,0)}) - 4 \alpha^2 (u_{(0,0)}^2 + u_{(1,0)} u_{(-1,0)}) - (\alpha^2-1)^2 \delta^2}{u_{(1,0)}-u_{(-1,0)}} \partial_{u_{(0,0)}}\,,\\
    & & \\
     {\mathbf{v}}_2 \,=\, \frac{2 \beta (\beta^2+1) u_{(0,0)} (u_{(0,1)}+u_{(0,-1)}) - 4 \beta^2 (u_{(0,0)}^2 +u_{(0,1)} u_{(0,-1)}) - (\beta^2-1)^2 \delta^2}{u_{(0,1)}-u_{(0,-1)}}\partial_{u_{(0,0)}}\,.
    \end{eqnarray*} \\
    
    \item {\bf{Q4}}
    
        {\em{Three-point generalized symmetries}} :
    \begin{eqnarray*} 
    {\mathbf{v}}_1 \,=\, \frac{(u_{(1,0)}-u_{(-1,0)}) f_{,u_{(1,0)}}(u_{(0,0)},u_{(1,0)},\alpha) - 2 f(u_{(0,0)},u_{(1,0)},\alpha)}{u_{(1,0)}-u_{(-1,0)}} \partial_{u_{(0,0)}}\,,\\
    & & \\
     {\mathbf{v}}_2 \,=\, \frac{(u_{(0,1)}-u_{(0,-1)}) f_{,u_{(0,1)}}(u_{(0,0)},u_{(0,1)},\beta) - 2 
f(u_{(0,0)},u_{(0,1)},\beta)}{u_{(0,1)}-u_{(0,-1)}}\partial_{u_{(0,0)}}\,,
    \end{eqnarray*}
     where
$$f(u_{(0,0)},u_{(1,0)},\alpha) \,=\,\left(u_{(0,0)} u_{(1,0)} + \alpha ( u_{(0,0)} + u_{(1,0)} ) + \frac{g_2}{4}\right)^2 - (u_{(0,0)}+u_{(1,0)}+\alpha) (4 \alpha u_{(0,0)} u_{(1,0)} - g_3)\,.$$
\end{itemize}

The extended symmetry transformations acting on the lattice parameters along with the corresponding determining equation have been presented in Section \ref{param}. Using these results, we find that the integrable lattice equations of Adler, Bobenko and Suris admit the extended symmetries of the following list.

\begin{itemize}

\item {\bf{H1}}

{\em{Point symmetries}} :  ${\mathbf{x}}_4 = u_{(0,0)} \partial_{u_{(0,0)}}  + 2 \alpha \partial_\alpha + 2 \beta \partial_\beta,\,\,{\mathbf{x}}_5 = \partial_\alpha + \partial_\beta\,.$ \\

{\em{Three-point generalized symmetries}} :
$$ {\mathbf{v}}_5 =  A(n) {\mbox{\bf{v}}}_1 + \left(A(n)-A(n+1)\right) \partial_\alpha ,\,\,{\mathbf{v}}_6 =  B(m) {\mbox{\bf{v}}}_3 + \left(B(m)-B(m+1)\right) \partial_\beta. $$ 

\item {\bf{H2}}

{\em{Point symmetries}} :  ${\mathbf{x}}_2 =  u_{(0,0)} \partial_{u_{(0,0)}}  +  \alpha \partial_\alpha + \beta \partial_\beta,\,\,{\mathbf{x}}_3 = \partial_{u_{(0,0)}} - 2 \partial_\alpha - 2 \partial_\beta \,.$ \\

{\em{Three-point generalized symmetries}} :
$$ {\mathbf{v}}_5 =  A(n) {\mbox{\bf{v}}}_1 + \left(A(n)-A(n+1)\right) \partial_\alpha ,\,\,{\mathbf{v}}_6 =  B(m) {\mbox{\bf{v}}}_3 + \left(B(m)-B(m+1)\right) \partial_\beta. $$ 

\item {\bf{H3}}

\begin{enumerate}
\item $\delta=0$.

{\em{Point symmetries}} :  ${\mathbf{x}}_3 = \alpha \partial_\alpha + \beta \partial_\beta \,.$ \\

{\em{Three-point generalized symmetries}} :
$$ {\mathbf{v}}_3 = A(n) {\mathbf{v}}_1\,-\,\left(A(n)-A(n+1)\right) \alpha \partial_\alpha,\,\,{\mathbf{v}}_4 = B(m) {\mathbf{v}}_2\,-\,\left(B(m)-B(m+1)\right) \beta \partial_\beta. $$ 

\item $\delta \neq 0$. 

{\em{Point symmetries}} :  ${\mathbf{x}}_2 = u_{(0,0)} \partial_{u_{(0,0)}} + 2 \alpha \partial_\alpha + 2 \beta \partial_\beta\,.$ \\

{\em{Three-point generalized symmetries}} :
$$ {\mathbf{v}}_3 = A(n) {\mathbf{v}}_1\,-\,\left(A(n)-A(n+1)\right) \alpha \partial_\alpha,\,\,{\mathbf{v}}_4 = B(m) {\mathbf{v}}_2\,-\,\left(B(m)-B(m+1)\right) \beta \partial_\beta. $$
\end{enumerate}

\item {\bf Q1}

\begin{enumerate}
\item $\delta=0$.

{\em{Point symmetries}} :  ${\mathbf{x}}_4 = \alpha \partial_\alpha + \beta \partial_\beta \,.$ \\

{\em{Three-point generalized symmetries}} :
$$ {\mathbf{v}}_3 = A(n) {\mathbf{v}}_1\,-\,\left(A(n)-A(n+1)\right) \alpha \partial_\alpha,\,\,{\mathbf{v}}_4 = B(m) {\mathbf{v}}_2\,-\,\left(B(m)-B(m+1)\right) \beta \partial_\beta. $$ 

\item $\delta \neq 0$. 

{\em{Point symmetries}} :  ${\mathbf{x}}_2 = u_{(0,0)} \partial_{u_{(0,0)}} + \alpha \partial_\alpha + \beta \partial_\beta.$ \\

{\em{Three-point generalized symmetries}} :
$$ {\mathbf{v}}_3 = A(n) {\mathbf{v}}_1\,-\,\left(A(n)-A(n+1)\right) \alpha \partial_\alpha,\,\,{\mathbf{v}}_4 = B(m) {\mathbf{v}}_2\,-\,\left(B(m)-B(m+1)\right) \beta \partial_\beta. $$ 
\end{enumerate}

\item {\bf Q2}

{\em{Point symmetries}} :  ${\mathbf{x}}_1 = 2 u_{(0,0)} \partial_{u_{(0,0)}} + \alpha \partial_\alpha + \beta \partial_\beta \,.$ \\

{\em{Three-point generalized symmetries}} :
$$ {\mathbf{v}}_3 = A(n) {\mathbf{v}}_1\,-\,\left(A(n)-A(n+1)\right) \alpha \partial_\alpha,\,\,{\mathbf{v}}_4 = B(m) {\mathbf{v}}_2\,-\,\left(B(m)-B(m+1)\right) \beta \partial_\beta. $$ 

\item {\bf Q3}

{\em{Three-point generalized symmetries}} :
\begin{eqnarray*}
{\mathbf{v}}_3 &=& A(n){\mathbf{v}}_1\,-\,2\,\left(A(n)-A(n+1)\right) \alpha^2 (\alpha^2-1) \partial_\alpha\,,\\
     {\mathbf{v}}_4 &=& B(m) {\mathbf{v}}_2\,-\,2\,\left(B(m)-B(m+1)\right) \beta^2 (\beta^2 -1) \partial_\beta \,.
\end{eqnarray*}

\item {\bf Q4}

{\em{Three-point generalized symmetries}} :
\begin{eqnarray*}
 {\mathbf{v}}_3 &=& A(n){\mathbf{v}}_1\,+\,\left(A(n)-A(n+1)\right) \left(4 \alpha^3 - g_2 \alpha - g_3 \right) \partial_\alpha\,,\\
     {\mathbf{v}}_4 &=& B(m) {\mathbf{v}}_2\,+\,\left(B(m)-B(m+1)\right) \left(4 \beta^3 - g_2 \beta - g_3 \right) \partial_\beta \,.
\end{eqnarray*} 
\end{itemize}

\section{Symmetry reductions to discrete analogues of the Painlev\'e equations} \label{red}

In the same way as in the case of the partial differential equations, the symmetries of a partial difference equation provide an effective means of constructing (whole classes of) special solutions. They are the ones that retain their form when acted on by some of the transformations leaving the equation invariant.  Hence, they are referred to as group invariant solutions. 

Group invariant solutions of certain integrable lattice equations are known to be associated with integrable mappings and discrete versions of the Painlev\'e equations. Reductions to integrable mappings first appeared in \cite{Pap1}, where the periodic boundary value problems for the lattice potential KdV were studied (see also \cite{cap, tp:Quisp2}). On the other hand, reductions to discrete Painlev\'e equations originally appeared in \cite{Pap2}, where the higher symmetries of the latter equation were exploited. For a more recent account on the subject we refer to \cite{tp:ramani, Hab, tp:Nij3, frank}. 

Here, we use the results obtained above to study group invariant solutions of the class of lattice equations introduced in Section \ref{sec3}, as well as initial value problems associated with them. More specifically, we relate the existence of symmetry reductions to specific Cauchy problems, leading to a unique solution. The existence of such group invariant solutions essentially follows from the fact that every equation in the class admits a five-point generalized symmetry with generator $\mathbf{v}_n+\lambda \mathbf{v}_m$ (see Proposition \ref{prop3p}). In the generic case, these reductions lead to four dimensional mappings. We show that the existence of a Lie point symmetry of the lattice equation, which is also compatible with the five-point symmetry constraint, suffices to reduce the resulting mapping to a third dimensional one. As an illustration we apply the results just mentioned to the lattice potential KdV {\bf{(H1)}}.

\subsection{Symmetry reductions and Cauchy problems}

Let $Q=0$ be a partial difference equation and $\mathbf{v}=R[u]\,\partial_{u_{(0,0)}}$ an infinitesimal generator of a symmetry transformation acting on the space of the dependent variable $u$, only. 
The square brackets in the symmetry characteristic $R[u]$ means that, the latter is, in general, a function of $(n,m,u_{(0,0)})$ and the shifted values of $u$ in both directions of the lattice up to some order $k$, i.e.
$R:\mathds{Z}^2 \times {\rm J}^{(k,-k)} \mapsto \mathds{C}$ for some fixed $k \in \mathds{N}$.

\begin{deff}
A function $u:\mathds{Z}^2 \mapsto \mathds{C}$ is called an invariant solution of
the lattice equation $Q=0$ under the symmetry $\mathbf{v}$,   
if it satisfies the lattice equation $Q=0$ and the compatible constraint $\mathbf{v}(u)=0$
(equivalently $R[u]=0$). 
\end{deff}

A natural question is whether non-trivial solutions of this kind do really exist. 
Consider a generic lattice equation $Q=0$ of the form (\ref{eq:genform}), where $Q$ satisfies the properties of Section \ref{sec3}. According to Proposition \ref{prop3p}, the lattice equation $Q=0$ always admits two generalized symmetry generators $\mathbf{v}_n$, $\mathbf{v}_m$. Thus, the invariant solutions under the symmetry generator
$$\mathbf{v}_{\mathcal{C}}\,=\,\mathbf{v}_n\,+\,\lambda\,\mathbf{v}_m\,$$
satisfy the lattice equation and the compatible symmetry constraint  
\begin{equation}
{\mathcal{C}}[u]:=  R(u_{(0,0)},u_{(1,0)},u_{(-1,0)};\alpha,\beta)\,+\,\lambda\,R(u_{(0,0)},u_{(0,1)},u_{(0,-1)};\beta,\alpha)=0, \label{symcon}
\end{equation}
where
$$R(u,x,y;\alpha,\beta)=\frac{h(u,x;\alpha,\beta)}{x-y} - \frac{1}{2}h_{,x}(u,x;\alpha,\beta)\,.$$

Let us now consider the following initial value problem. Given the values of $u$ assigned on the four points depicted with black in Figure \ref{fig:BSQ}, one can determine uniquely the values of $u$ along three vertical columns of the square grid, by using the lattice equation ($\raisebox{3pt}{$\!\!\phantom{i}_{\bigcirc}$}$ points) and the symmetry
constraint ($+$ points). It should be noted that, due to the affine linearity of the function $Q$, the symmetry constraint (\ref{symcon} can be solved uniquely for each one of the values  $u_{(-1,0)}$, $u_{(1,0)}$, $u_{(0,-1)}$ and $u_{(0,1)}$. Then, the values of $u$ on the remaining vertices of the lattice 
can be determined, by using successively the symmetry constraint. In this setting there will be points where
the values  of $u$ can be determined by using either the lattice equation, or the symmetry constraint
(\raisebox{3pt}{$\!\!\phantom{i}_{\bigoplus}$}  points). However, since the symmetry constraint is by definition compatible with the lattice equation, the values determined by the two different ways described above are, necessarily, identical. Thus, for generic initial data, such a group-invariant solution exists and is unique 
(see the discussion in \cite{vsesa, Pap2, frank}). Let it be noted however, that such a symmetry reduction leads to four-dimensional mappings.

\begin{figure}[h]
\begin{minipage}{18pc}
\centertexdraw{
\setunitscale 0.5
\linewd 0.02 \arrowheadtype t:F 

\move (0. -1.)  
\lvec (1 -1.) 
\lvec (2. -1) \move (1 -1)
\lvec (1 0) 

\linewd 0.01
\lpatt(0.12 0.12)
\move (0. -2.)  
\lvec (1 -2.) 
\lvec (2. -2) \move (1 -1.85)
\lvec (1 -1) 

\linewd 0.01
\lpatt(0.12 0.12)
\move (0. 0.)  
\lvec (1 0.) 
\lvec (2. 0) \move (1 0.)
\lvec (1 0.875) 

\lpatt() 
\linewd 0.02

\move (2.925 0) \lvec (3.075 0) 
\move (3 -0.075) \lvec (3 0.075)

\move (1 0.925) \lvec (1 1.075) 
\move (0.925 1) \lvec (1.075 1)

\move (2 1) \fcir f:1.0 r:0.05 \move (2 1) \lcir r:0.05
\move (2 0) \fcir f:1.0 r:0.05 \move (2 0) \lcir r:0.05
\move (2 -2.) \fcir f:1.0 r:0.05 \move (2 -2.) \lcir r:0.05 

\move (1 -1.925) \lvec (1 -2.075) 
\move (0.925 -2.) \lvec (1.075 -2.)

\move (0 -1) \fcir f:0.0 r:0.075
\move (0 -2) \fcir f:1.0 r:0.05 \move (0 -2) \lcir r:0.05
\move (0 1) \fcir f:1.0 r:0.05 \move (0 1) \lcir r:0.05

\move (0. 0.) \fcir f:1.0 r:0.05 \move (0. 0.) \lcir r:0.05
\move (1 0.) \fcir f:0.0 r:0.075
\move (1 -1.) \fcir f:0.0 r:0.075
\move (2. -1) \fcir f:0.0 r:0.075

\linewd 0.015
\move (2.925 -1) \lvec (3.075 -1) 
\move (3 -1.075) \lvec (3 -0.925)
\move (3 -1) \lcir r:0.075
}
\caption{A Cauchy problem on the lattice.} \label{fig:BSQ}
\end{minipage}\hspace{1.pc}%
\begin{minipage}{18pc}
\centertexdraw{
\setunitscale 0.5
\linewd 0.00 \arrowheadtype t:F 

\move (0 2.0) \htext{\phantom{.}} 

\move (1.5 1.5)
\lvec (2. 1) \lvec (2. 1.5) \lvec (1.5 1.5) \ifill f:0.8

\move (2. 1.)
\lvec (2. 0) \lvec (1.5 0.5) \lvec (2.0 1.) \ifill f:0.8

\move (2. 0.)
\lvec (2. -1) \lvec (1.5 -0.5) \lvec (2.0 0.) \ifill f:0.8

\move (2. -1.5)
\lvec (2. -1) \lvec (1.5 -1.5) \lvec (2.0 -1.5) \ifill f:0.8

\move (0.5 1.5) 
\lvec (1.5 1.5) \lvec (1.0 1.) \lvec (0.5 1.5) \ifill f:0.8

\move (0.5 -1.5) 
\lvec (1.5 -1.5) \lvec (1.0 -1.) \lvec (0.5 -1.5) \ifill f:0.8

\move (-0.5 -1.5) 
\lvec (-1.0 -1.5) \lvec (-1.0 -1.) \lvec (-0.5 -1.5) \ifill f:0.8

\move (-0.5 1.5) 
\lvec (-1.0 1.5) \lvec (-1.0 1.) \lvec (-0.5 1.5) \ifill f:0.8

\move (0. 1.) 
\lvec (0.5 1.5) \lvec (-0.5 1.5) \lvec (0. 1.) \ifill f:0.8

\move (-1. 0.) 
\lvec (-0.5 0.5) \lvec (-1 1) \lvec (-1. 0.) \ifill f:0.8

\move (-1. 0.) 
\lvec (-0.5 -0.5) \lvec (-1 -1) \lvec (-1. 0.) \ifill f:0.8

\move (0. -1.) 
\lvec (-0.5 -1.5) \lvec (0.5 -1.5) \lvec (0. -1.) \ifill f:0.8

\move (0.5 -0.5) 
\lvec (1. -1.0) \lvec (1.5 -0.5) \lvec (1. 0.0)  \lvec (0.5 -0.5) \ifill f:0.8

\move (1. 0.) 
\lvec (1.5 0.5) \lvec (1 1) \lvec (0.5 0.5) \lvec (1. 0.) \ifill f:0.8

\linewd 0.02 

\move (1.5 -1.5)
\lvec (1. -1.) \lvec (1.5 -0.5) \lvec (1 0) \lvec (1.5 0.5) \lvec (1 1) \lvec (1.5 1.5)
\move (1.5 -1.5)
\lvec (2.0 -1.) \lvec (1.5 -0.5) \lvec (2 0) \lvec (1.5 0.5) \lvec (2 1) \lvec (1.5 1.5)
\move (0. 0.)  
\lvec (0.5 0.5) \lvec (0 1) \lvec (-0.5 0.5) \lvec (0. 0.) \ifill f:0.8
\move (0. 0.)  
\lvec (0.5 0.5) \lvec (0 1) \lvec (-0.5 0.5) \lvec (0. 0.)

\move (0. 0.)  
\lvec (-0.5 -0.5) \lvec (0 -1) \lvec (0.5 -0.5) \lvec (0. 0.) \ifill f:0.8
\move (0. 0.)  
\lvec (-0.5 -0.5) \lvec (0 -1) \lvec (0.5 -0.5) \lvec (0. 0.) 

\move (0. -1.0) 
\lvec (0.5 -1.5) \lvec (1. -1) \lvec (0.5 -0.5) \lvec (1. 0.) \lvec (0.5 0.5) \lvec (1. 1.) \lvec (0.5 1.5) \lvec (0. 1.)

\move (0. -1.0) 
\lvec (-0.5 -1.5) \lvec (-1. -1) \lvec (-0.5 -0.5) \lvec (-1. 0.) \lvec (-0.5 0.5) \lvec (-1. 1.) \lvec (-0.5 1.5) \lvec (0. 1.)


\move (-1. -0.5) \fcir f:1.0 r:0.075
\move (0 0.5) \fcir f:1.0 r:0.075
\move (0 -0.5) \fcir f:1.0 r:0.075
\move (1. -0.5) \fcir f:1.0 r:0.075
\move (-1. -0.5) \lcir r:0.075
\move (0 0.5) \lcir r:0.075
\move (0 -0.5) \lcir r:0.075
\move (1. -0.5) \lcir r:0.075

\move (1 0) \fcir f:0.0 r:0.075
\move (1.5 -0.5) \fcir f:0.0 r:0.075

\move (-0.5 -0.5) \fcir f:0.0 r:0.075
\move (0. 0.) \fcir f:0.0 r:0.075
\move (0.5 -0.5) \fcir f:0.0 r:0.075

\begin{scriptsize} 
\move (-0.6 -0.85) \htext{${x_1}$}
\move (-0.12 0.13) \htext{${x_2}$}
\move (0.37 -0.9) \htext{$x_1^{'}$}
\move (0.4 -0.35) \htext{${x_3}$}
\move (1.4 -0.35) \htext{$x_3^{'}$}
\move (0.88 0.13) \htext{$x_2^{'}$}
\end{scriptsize}
}
\caption{The Cauchy problem on the dual (chessboard) lattice.} \label{fig:Q}
\end{minipage} 
\end{figure}

Nevertheless, under certain conditions, symmetry reductions of Cauchy problem of the type under consideration lead to mappings which are three-dimensional. To see how this is obtained, let us require that a given Lie point symmetry of the lattice equation, is also a symmetry of the constraint $\mathcal{C}[u]=0$. If $\mathbf{x}=X(n,m,u_{(0,0)})\, \partial_{u_{(0,0)}}$ is the generator of the Lie point symmetry, then the above requirement is equivalent to the following commutation relation\footnote{Indeed, the commutation relation $\left[\mathbf{x} , \mathbf{v}_{\mathcal{C}} \,\right] =0$, yields $\mathbf{x} (\mathcal{C}[u]) - 
\mathcal{C}[u] \partial_{u_{(0,0)}} ( X(n,m,u_{(0,0)}) )=0$ which, in virtue of $\mathcal{C}[u]=0$, means that 
$\mathbf{x}$ is also a symmetry generator of the constraint $\mathcal{C}[u]=0$.}
\begin{equation}
\left[\mathbf{x} ,  \mathbf{v}_{\mathcal{C}} \,\right] =0, 
\quad {\rm mod} \quad \mathcal{C}[u] =0 \,. \label{eq:comcon}
\end{equation}

The existence of the point symmetry generated by $\mathbf{x}$ allows one 
to define lattice invariants ({\em reduced variables}) assigned on the edges of 
the square grid, in terms of which, the lattice equation can be casted as a map. 
The dynamics of this map is restricted to the white squares of the chessboard in Figure \ref{fig:Q}.
Moreover, since $\mathbf{x}$ is also a symmetry of the constraint $\mathcal{C}[u]=0$ 
(defined now on the vertices of the shaded squares of the chessboard), 
the latter can be written in terms of the invariants of the 
vector field $\mathbf{x}$ on $\rm{J}^{(1,-1)}$.
In this way, the corresponding invariant solutions are constructed from the 
associated Cauchy problem with initial values $x_1,x_2,x_3$ assigned on the 
edges (Figure \ref{fig:Q}), and, consequently, the mappings obtained by this procedure are, 
in general, three dimensional. 

A specific example illustrating the case of reductions to three-dimensional mappings is provided by the lattice 
potential KdV equation (\ref{H1}). The most general symmetry constraint on a cross configuration of points is 
obtained from a linear combination of the symmetry generators presented in the previous section, i.e.
\begin{equation}
{\mathbf{v}}_{\mathcal{C}} =  \sum_{i=1}^{3} \lambda_i\, {\mathbf{x}}_i + 
\sum_{i=1}^{4} \mu_i \, {\mathbf{v}}_i\,, \label{eq:kdvcon}
\end{equation}
where $\lambda_i, \mu_i$ are arbitrary complex parameters.
Here, $\left\lbrace \mathbf{x}_1,\mathbf{x}_2,\mathbf{x}_3\right\rbrace$ is 
the set of Lie point symmetry generators, which span a Lie algebra $\mathfrak{g}$ 
isomorphic to ${\mathfrak{so}}(1,1)$, and 
$\left\lbrace \mathbf{v}_1,\mathbf{v}_2,\mathbf{v}_3,\mathbf{v}_4 \right\rbrace$ 
is the set of the three-point generalized symmetries .
We note that, under the transformation $u \mapsto (-1)^{n+m} u$, the symmetry 
generator $\mathbf{x}_2$ is mapped to $\mathbf{x}_1$. 
Thus, for the symmetry reduction using the invariants of one-dimensional subalgebras 
of $\mathfrak{g}$, it is sufficient to consider two inequivalent cases, namely 
$\mathbf{x}=\mathbf{x}_1$ or $\mathbf{x}_3$. 

\subsection{Symmetry reduction of $\,${\bf{H1}} using the invariants of $\,\mathbf{x}_1$}

We consider first the case $\mathbf{x}=\mathbf{x}_1$ and recast the lattice equation
$\mathbf{H1}$ as an invertible map 
$\mathcal{R}:\mathds{CP}^1\times \mathds{CP}^1 \mapsto \mathds{CP}^1\times \mathds{CP}^1$.
To this end, we introduce the following lattice invariants along the orbits of $\mathbf{x}_1$
\begin{equation}
v_{(0,0)}=u_{(1,0)}-u_{(0,0)},\, w_{(0,0)}=u_{(1,1)} - u_{(1,0)},
\, v_{(0,1)}=u_{(1,1)}-u_{(0,1)}, \,
w_{(-1,0)}=u_{(0,1)}-u_{(0,0)}, \label{eq:invx1}
\end{equation}
where $v_{(i,j)},w_{(i,j)} \in \rm{J}^{(2)}_{\ast}$ and
$\rm{J}^{(2)}_{\ast}$ is the space obtained from $\rm{J}^{(2)}$ by removing 
$(u_{(2,0)},u_{(0,2)})$.

\begin{figure}[h]
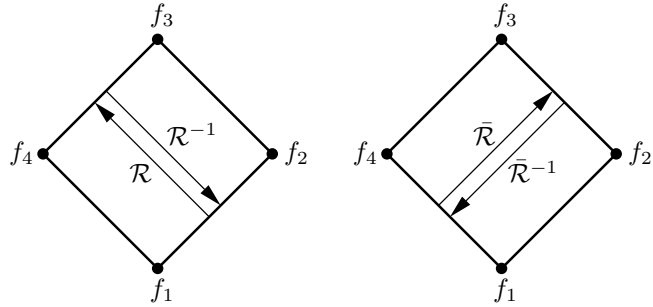

\centertexdraw{
\setunitscale 0.6
\linewd 0.02 \arrowheadtype t:F \arrowheadsize l:0.2 w:0.1
\move (0. -1.)  
\lvec (1. 0.) 
\lvec (0 1.) 
\lvec (-1. 0.) 
\lvec (0. -1)
\move (0 -1) \fcir f:0.0 r:0.05
\move (0 1) \fcir f:0.0 r:0.05
\move (-1 0) \fcir f:0.0 r:0.05
\move (1 0) \fcir f:0.0 r:0.05

\linewd 0.01
\move (0.45 -0.55)   \avec (-0.55 0.45)
\move (-0.45 0.55)   \avec  (0.55 -0.45)

\linewd 0.02
\move (3. -1.)  
\lvec (4. 0.) 
\lvec (3 1.) 
\lvec (2. 0.) 
\lvec (3. -1)
\move (3 -1) \fcir f:0.0 r:0.05
\move (3 1) \fcir f:0.0 r:0.05
\move (2 0) \fcir f:0.0 r:0.05
\move (4 0) \fcir f:0.0 r:0.05

\linewd 0.01

\move (3.55 0.45)  \avec (2.55 -0.55) 

\move (2.45 -0.45)   \avec (3.45 0.55) 

\move (0.08 0.08) \htext {$\mathcal{R}^{-1}$}
\move (-0.25 -0.25) \htext {$\mathcal{R}$}
\move (2.75 0.08) \htext {$\bar{\mathcal{R}}$}
\move (3.05 -0.25) \htext {$\bar{\mathcal{R}}^{-1}$}

\move (-0.075 -1.3) \htext {$f_1$}
\move (1.1 -0.1) \htext {$f_2$}
\move (-0.075 1.1) \htext {$f_3$}
\move (-1.3 -0.1) \htext {$f_4$}

\move (2.925 -1.3) \htext {$f_1$}
\move (4.1 -0.1) \htext {$f_2$}
\move (2.925 1.1) \htext {$f_3$}
\move (1.7 -0.1) \htext {$f_4$}

\move (-1.2 1.2) \htext{\phantom{o}}
}
\caption{The invertible maps $\mathcal{R}$, $\bar{\mathcal{R}}$. 
Their dynamics is restricted to the white squares of the chessboard.} \label{fig:quadri}
\end{figure}

There is a functional relation among the above invariants, namely
\begin{equation}
v_{(0,0)} + w_{(0,0)}= v_{(0,1)} + w_{(0,-1)}\,, \label{eq:invx11}
\end{equation}
following from the fact that the space of invariants along the orbits of $\mathbf{x}$  
on $\rm{J}^{(1)}_{\ast}$ is three dimensional. 
On the other hand, since $\mathbf{x}$ is a symmetry generator of equation {\bf H1}, 
the latter can be written in terms of the invariants (\ref{eq:invx1}) in the form 
\begin{equation}
(v_{(0,0)}+w_{(0,0)})(v_{(0,0)}-w_{(-1,0)})+r=0 \,, \label{eq:invH1}
\end{equation}
where $r=\alpha-\beta$. 
Equations (\ref{eq:invx11}), (\ref{eq:invH1}) can be {\em uniquely} solved for
$v_{(0,1)}$, $w_{(-1,0)}$ in terms of $v_{(0,0)}$, $w_{(0,0)}$, and conversely, 
implying the existence of the following invertible map
\begin{equation}
\mathcal{R}(f_1,f_2) = (f_3,f_4) = \left( f_2 - \frac{r}{f_1+f_2}\,,
f_1 + \frac{r}{f_1+f_2}\right) \,,
\label{eq:YB1mapH1}
\end{equation}
where 
\begin{equation}
(f_1,f_2,f_3,f_4) = (v_{(0,0)},w_{(0,0)},v_{(0,1)},w_{(-1,0)})\,.
\end{equation}
Moreover, equations (\ref{eq:invx11}), (\ref{eq:invH1}) can be {\em uniquely} solved 
for $v_{(0,1)}$, $w_{(0,0)}$ in terms of $v_{(0,0)}$, $w_{(-1,0)}$, 
implying also the existence of the invertible map
\begin{equation}
\bar{\mathcal{R}}(f_1, f_4) = (f_3, f_2) =
\left( -f_4 + \frac{r}{f_4-f_1} \,, -f_1 + \frac{r}{f_4-f_1} \right) \,.
\label{eq:YB2mapH1}
\end{equation}
The existence of the above invertible map $\bar{\mathcal{R}}$ 
is associated with the {\em quadrarationality} property of the birational map $\mathcal{R}$.
Maps with this property were studied recently in \cite{ABS2},
in connection with the Yang-Baxter relation. Here, the quadrarationality property of the map $R$
allows one to uniquely determine two of the four values $f_i$ assigned on the vertices of the 
white squares of the chessboard in terms of the remaining two, as shown in Figure \ref{fig:quadri}.

We proceed by writing the symmetry constraint $\mathbf{v}_{\mathcal{C}}(u)=0$, 
where $\mathbf{v}_{\mathcal{C}}$
is given by equation (\ref{eq:kdvcon}), in terms of the invariants (\ref{eq:invx1}) and their shifts.
To this end, it is necessary that relation (\ref{eq:comcon}) holds. 
A direct calculation shows that
\begin{equation}
\left[\mathbf{x}_1 ,  \mathbf{v}_{\mathcal{C}} \,\right] = \lambda_3 \, \mathbf{x}_2 +
\frac{1}{2}\,\frac{\mu_2-\mu_4}{\alpha-\beta} \, \mathbf{x}_1,  \label{eq:comconx1vc}
\end{equation}
from which we conclude that the parameters $\lambda_i$, $\mu_i$ should be rectricted by the following relations
\begin{equation}
\lambda_3=0\,,\quad \mu_2=\mu_4\,.
\end{equation}
Taking into account the above relations, one can write the symmetry constraint $\mathbf{v}_{\mathcal{C}}(u)=0$ 
in terms of the invariants (\ref{eq:invx1}) in the form
\begin{equation}
\mathcal{C}(n,m,g_1,g_2,g_3,g_4):=
\lambda_1 + \lambda_2\,(-1)^{n+m} + \frac{\mu_2 \, n + \mu_1}{g_1+g_2} + 
\frac{\mu_2 \, m + \mu_3}{g_3+g_4} =0\,. \label{eq:conKdV} 
\end{equation}
where 
\begin{equation}
(g_1,g_2,g_3,g_4)=(v_{(0,0)},v_{(-1,0)},w_{(-1,0)},w_{(-1,-1)})\,.
\end{equation}

We consider now the Cauchy problem on the chessboard with initial values $(x_1,x_2,x_3)$, as 
shown in Figure \ref{fig:Q}. The updated values $(x_1^{\prime},x_2^{\prime},x_3^{\prime})$ in the $n$-direction of the lattice 
can be found by using the invertible maps $\mathcal{R}$, $\bar{\mathcal{R}}$ and the 
symmetry constraint (\ref{eq:conKdV}). Indeed, the value $x_2^{\prime}$ is the second component 
of the map $\bar{\mathcal{R}}(x_3,x_2)$, and obviously $x_3^{\prime}=x_1$. The updated value
$x_3^{\prime}$ is computed in three steps. First, we find $x_4$ by solving the equation 
$\mathcal{C}(n,m,x_3,x_1,x_4,x_2)=0$ for $x_4$, next from the map $R^{-1}(x_3,x_4)=(x_5,x_6)$ 
we pick the second component $x_6$, and finally the value $x_3^{\prime}$ is found by solving the
equation $\mathcal{C}(n+1,m,x_3^{\prime},x_3,x_6,x_2^{\prime})=0$, for the corresponding variable. 
A straightforward calculation shows that the updated values $(x_1^{\prime},x_2^{\prime},x_3^{\prime})$
are found from the following non-autonomous system of first order ordinary difference equation in the variable $n$,
\begin{eqnarray}
 x_1^{\prime} &=& x_3\,,\\
 x_2^{\prime} &=& -x_3 + \frac{r}{x_3-x_2}\,, \\
 x_3^{\prime} &=& -x_3 + \frac{r\, c(n+1,\mu_1)}{\left(\frac{c(n,\mu_1)}{x_1+x_3} + \lambda(n+m)\right)(x_2-x_3)^2 +  c(m,\mu_3)\,(x_2-x_3) - r\lambda(n+m+1) },
\end{eqnarray}
where $c(n,\mu)=\mu_2\,n + \mu$ and $\lambda(n)=\lambda_1+\lambda_2\,(-1)^{n}$. 
The above system can be decoupled for the variable $y(n):=x_2^{\prime}+x_3=x_2(n+1,m)+x_3(n,m)$ 
leading to the second order difference equation
\begin{equation}
 \frac{r\,c(n+1,\mu_1)}{y(n+1)\,y(n) +r} + \frac{r\,c(n,\mu_1)}{y(n)\,y(n-1) +r} =
c(n+1,\mu_1) + c(m,\mu_3) + y(n)\,\lambda(n+m+1) - \frac{r}{y(n)}\,\lambda(n+m),
\end{equation}
which is known as the asymmetric, alternate discrete Painlev\'e II equation  \cite{FGR}, \cite{vas}.

\subsection{Symmetry reduction of $\,${\bf{H1}} using the invariants of $\,\mathbf{x}_3$}

We now turn to the case where we use the invariants of $\mathbf{x}=\mathbf{x}_3$. Following the preceding considerations 
(see also \cite{Pap3}), we introduce the following invariants on $\rm{J}^{(2)}_{\ast}$ along the orbits of $\mathbf{x}_3$ 
\begin{equation}
 v_{(0,0)}=u_{(1,0)}\,u_{(0,0)},\quad w_{(0,0)}=u_{(1,1)} \, u_{(1,0)},
\quad v_{(0,1)}=u_{(1,1)}\,u_{(0,1)}, \quad
w_{(-1,0)}=u_{(0,1)}\,u_{(0,0)}\,. \label{eq:invx3}
\end{equation}
The above invariants are functionally related by
\begin{equation}
v_{(0,0)}\,v_{(0,1)}  = w_{(0,0)}\, w_{(0,-1)}\,. \label{eq:relinvx3}
\end{equation}
On the other hand, equation {\bf H1} can be written in terms of the invariants (\ref{eq:invx3}) in the form 
\begin{equation}
v_{(0,0)} - w_{(-1,0)} - w_{(0,0)} + v_{(0,0)}=r \,. \label{eq:inv2H1}
\end{equation}
Using equations (\ref{eq:relinvx3}), (\ref{eq:inv2H1}) we obtain the invertible maps 
$\mathcal{R}:(f_1,f_2)\mapsto (f_3,f_4)$,  $\bar{\mathcal{R}}(f_1,f_4)\mapsto (f_3,f_2)$. 
Explicitly they are defined by
\begin{subequations}
\begin{eqnarray}
\mathcal{R}(f_1,f_2) &=& (f_3,f_4) = \left( f_2\, \big( 1 + \frac{r}{f_2-f_1} \big)\,,
f_1\, \big( 1 + \frac{r}{f_2-f_1}\big) \right) \,,\\
{\bar{\mathcal{R}}}(f_1,f_4) &=& {\mathcal{R}}(f_1,f_4) \,=\,(f_3,f_2)\,,
\end{eqnarray}
\label{eq:YB3mapH1}
\end{subequations}
where $(f_1,f_2,f_3,f_4) = (v_{(0,0)},w_{(0,0)},v_{(0,1)},w_{(-1,0)})$. Next we 
write the symmetry constraint $\mathbf{v}_{\mathcal{C}}(u)=0$ in terms of the invariants (\ref{eq:invx3}), where $\mathbf{v}_{\mathcal{C}}$ is given by (\ref{eq:kdvcon}). 
A straightforward calculation leads to the following result
\begin{equation}
\left[\mathbf{x}_3 , \mathbf{v}_{\mathcal{C}} \,\right] = -\lambda_1 \, \mathbf{x}_2 -
\lambda_2 \, \mathbf{x}_1 \,.  \label{eq:comconx3vc}
\end{equation}
Hence, relation (\ref{eq:comcon}) holds whenever 
\begin{equation}
\lambda_1=0\,,\quad \lambda_2=0\,.
\end{equation}
In this setting, the symmetry constraint $\mathbf{v}_{\mathcal{C}}(u)=0$ 
is written in terms of the invariants (\ref{eq:invx3}) as follows 
\begin{equation}
\mathcal{C}(n,m,g_1,g_2,g_3,g_4):=
\lambda_3\,(-1)^{n+m} + \frac{\mu_2-\mu_4}{2 r}+ \frac{\mu_2 \, n + \mu_1}{g_1 - g_2} + 
\frac{\mu_4 \, m + \mu_3}{g_3 - g_4} =0\,, \label{eq:con2KdV} 
\end{equation}
where 
\begin{equation}
(g_1,g_2,g_3,g_4)=(v_{(0,0)},v_{(-1,0)},w_{(-1,0)},w_{(-1,-1)})\,.
\end{equation}

The corresponding Cauchy problem on the chessboard is reduced now to the solution 
of the ordinary difference equations defined by the mapping $M: (x_1,x_2,x_3) \mapsto  (x_1^{\prime},x_2^{\prime},x_3^{\prime})$. 
The updated values are found in the similar manner as in the previous case where now
the invertible maps $\mathcal{R}$, $\bar{\mathcal{R}}$ and the symmetry constraint are given 
by equations (\ref{eq:YB3mapH1}), (\ref{eq:con2KdV}), respectively. More precisely we have
\begin{eqnarray}
x_1^{\prime} &=& x_3 \,,\label{eq:Pa1}\\
x_2^{\prime} &=& x_3 \, \left(1+\frac{r}{x_2-x_3} \right) \,,\label{eq:Pa2}
\end{eqnarray}
and $x_3^{\prime}$ is found by solving 
\begin{equation}
 c(n+1) \, \frac{x_3^{\prime}-x_2^{\prime}}{x_3^{\prime}-x_1^{\prime}} + 
c(n) \, \frac{x_3-x_2}{x_3-x_1} = c(n+1) + d(m) +\lambda(n+m)(x_2-x_3) + \lambda(n+m+1)(x_2^{\prime}-x_3),
\label{eq:Pa3}
\end{equation}
where 
\begin{equation}
c(n)=\mu_2 n + \mu_1\,,\quad d(m)=\mu_4 m + \mu_3\,, \quad \lambda(n) = \lambda_3 (-1)^{n} +\frac{\mu_2-\mu_4}{2 r}.
\end{equation}
If $\mu_2=\mu_4$ then equation (\ref{eq:Pa3}) can be integrated once. Due to a compatibility condition,
the arbitrary function of $m$ in the discrete integration is specified up to a constant, yielding the following result
\begin{equation}
c(n) \, \frac{x_3-x_2}{x_3-x_1} = (-1)^{n+m} \, (\lambda_3 \, x_2 + \rho) + \frac{1}{2}(c(n)+d(m)) + \frac{\mu_2}{4}\,,
\label{eq:Pa4}
\end{equation}
where $\rho$ is the complex constant of integration. One may solve equation (\ref{eq:Pa4}) for $x_2$ and use the result to eliminate $x_2$ from equation (\ref{eq:Pa2}), using (\ref{eq:Pa1}). This leads to a second order difference equation for $x_1$, which we omit because of its length.

\section{Conclusions and perspectives} \label{concl}

We have presented a symmetry analysis of a class of lattice equations on $\mathds{Z}^2$, which are characterized by affine linearity and $\mathrm{D}_4$-symmetry. Once a specific equation in the class is given, the results summarized in Propositions \ref{prop3p} and \ref{prop5p} explicitly determine the characteristics of its three- and five-point generalized symmetry generators. Applied to the integrable equations obtained in classification \cite{ABS} these results allowed us to determine all Lie point, three- and five-point generalized symmetries admitted by the above equations. The results obtained in this fashion provide a proof that these lists of symmetries are exhaustive. From this point of view, the present work constitutes a generalization of the studies on these equations presented in \cite{Levi2}, \cite{hydol}, \cite{brew}, \cite{tp:TTP}.

The effectiveness of using Lie point symmetries of integrable lattice equations in obtaining Yang-Baxter maps was demonstrated in \cite{Pap4}, \cite{Pap3}. In particular, it was shown there that the Yang-Baxter variables can be chosen as invariants of the multi-parameter symmetry groups of the integrable lattice equations. Here, it was shown that this connection, combined with the quadrirationality property of the associated Yang-Baxter maps \cite{ABS2}, can be used in obtaining group invariant solutions of a lattice equation. Specifically, we considered a Cauchy problem for the lattice potential KdV equation compatible with the most general symmetry constraint on five points. The solution of this initial value problem was constructed by solving a second order mapping, which represents a discrete analogue of the Painlev\'e equations.

The preceding symmetry analysis may also be applied to those equations in the class considered above which do not have the consistency property. It would be interesting to consider the corresponding symmetry reductions and investigate various properties of the resulting mappings, such as the singularity confinement \cite{Pap1} and the algebraic entropy \cite{bv}. Moreover, it would be interesting to extend the results obtained here to lattice equations on $\mathds{Z}^2$, which possess symmetries other than the symmetries of the square. Work in this direction is in progress. 

\section*{Acknowledgments}
This work was supported by the research grant Pythagoras B-365-015 of the European Social Fund (EPEAEK II). The authors express their gratitude to V G Papageorgiou for fruitful discussions and comments on this work. Especially, for pointing out the possibility of expressing the symmetry reductions as Cauchy problems on the chessboard.

\section*{Appendix. Proof of Proposition \ref{prop3p}}

\appendix 
\setcounter{section}{1}

Let   
$$
{\mbox{\bf{v}}}^{(2)} \,=\, \sum_{i=0}^{1} \sum_{j=0}^{1} R(n+i,m+j,u_{(i,j)},u_{(i+1,j)},u_{(i-1,j)};\alpha,\beta) \partial_{u_{(i,j)}} \,
$$
be the second prolongation of the generator ${\mbox{\bf{v}}}$ of a three-point generalized symmetry. In this case, the infinitesimal symmetry criterion takes the form
\begin{eqnarray}
 Q_{,u_{(0,1)}} R(n,m+1,u_{(0,1)},u_{(1,1)},u_{(-1,1)};\alpha,\beta)
+  Q_{,u_{(1,1)}} R(n+1,m+1,u_{(1,1)},u_{(2,1)},u_{(0,1)};\alpha,\beta) \nonumber \\
 + Q_{,u_{(0,0)}} R(n,m,u_{(0,0)},u_{(1,0)},u_{(-1,0)};\alpha,\beta) + 
Q_{,u_{(1,0)}} R(n+1,m,u_{(1,0)},u_{(2,0)},u_{(0,0)};\alpha,\beta)  = 0\,.  \label{eq:DEq13point} 
\end{eqnarray}
The last equation must hold on every solution of the equation 
$$Q\,=\,Q(u_{(0,0)},u_{(1,0)},u_{(0,1)},u_{(1,1)};\alpha,\beta) \, = \,0\,,$$
and its shifted consequences, i.e.
$${\undertilde{Q}}\,=\,Q(u_{(-1,0)},u_{(0,0)},u_{(-1,1)},u_{(0,1)};\alpha,\beta) \,=\,0 \,, \,\,\,{\wtilde{Q}}\,=\,Q(u_{(1,0)},u_{(2,0)},u_{(1,1)},u_{(2,1)};\alpha,\beta) \,=\,0\,. $$
Using the above equations, one may express the values $u_{(2,1)}$, $u_{(0,0)}$ and $u_{(-1,0)}$ in terms of the remaining ones. 

One can eliminate the shifts of the characteristic $R$ from equation (\ref{eq:DEq13point}) by differentiating the latter w.r.t. $u_{(-1,1)}$, which yields
\begin{eqnarray} Q_{,u_{(0,1)}}\, R_{,u_{(-1,1)}}(n,m+1,u_{(0,1)},u_{(1,1)},u_{(-1,1)};\alpha,\beta) \nonumber \\
+\, Q_{,u_{(0,0)}}\, R_{,u_{(0,-1)}}(n,m,u_{(0,0)},u_{(1,0)},u_{(-1,0)};\alpha,\beta) \,\frac{\partial u_{(-1,0)}}{\partial u_{(-1,1)}}\,= 0 \,,  
\label{eq:DEq23point} \end{eqnarray}
since only $u_{(-1,0)}$ implicitly depends on $u_{(-1,1)}$ through the equation $\undertilde{Q} = 0$. Next, we divide equation (\ref{eq:DEq23point}) by $Q_{,u_{(0,1)}}$ and take the total derivative of the resulting equation w.r.t. $u_{(1,0)}$, i.e.
\begin{equation} \mbox{D}_{u_{(1,0)}} \left(\frac{Q_{,{u_{(0,0)}}}}{Q_{,{u_{(0,1)}}}} R_{,u_{(0,-1)}}(n,m,u_{(0,0)},u_{(1,0)},u_{(-1,0)};\alpha,\beta) \,\frac{\partial u_{(-1,0)}}{\partial u_{(-1,1)}} \right) = 0 \,, \label{eq:DEq33point} 
\end{equation}
where
$$ \mbox{D}_{u_{(1,0)}}\, = \,\partial_{u_{(1,0)}}\, +\, \frac{\partial u_{(0,0)}}{\partial u_{(1,0)}}\,\partial_{u_{(0,0)}}\, +\, \frac{\partial u_{(-1,0)}}{\partial u_{(0,0)}} \,\frac{\partial u_{(0,0)}}{\partial u_{(1,0)}}\,\partial_{u_{(0,0)}} \,. $$
\begin{figure}[h]
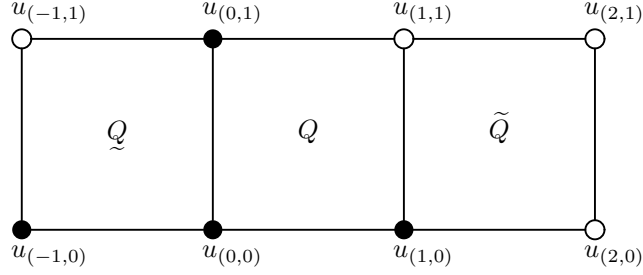

\centertexdraw{ \setunitscale 0.5
\linewd 0.02 \arrowheadtype t:F 
\htext(0 0.5) {\phantom{T}}
\move (-2.9 -2) \lvec (-1.1 -2)
\move (-0.9 -2) \lvec (2.9 -2) 
\move(-2.9 0) \lvec (0.9 0) \move(1.1 0) \lvec (2.9 0) \move(3 -.1) \lvec(3 -1.9)
\move (-3 -1.9) \lvec (-3 -.1)
\move (-3 -2) \fcir f:0.0 r:0.1 \move (-1 -2) \fcir f:0.0 r:0.1 \move (1 -2) \fcir f:0.0 r:0.1 \move (3 -2) \lcir r:0.1
\move (-3 0) \lcir r:0.1 \move (-1 0) \fcir f:0.0 r:0.1 \move (1 0) \lcir r:0.1 \move (3 0) \lcir r:0.1 
\move (-1 -1.9) \lvec (-1 0) \move (1 -2) \lvec (1 -.1) 
\htext (-3.1 -2.4) {$u_{(-1,0)}$} \htext (-1.1 -2.4) {$u_{(0,0)}$} \htext (.9 -2.4) {$u_{(1,0)}$} \htext (2.9 -2.4) {$u_{(2,0)}$}
\htext (-3.1 .15) {$u_{(-1,1)}$} \htext (-1.1 .15) {$u_{(0,1)}$} \htext (.9 .15) {$u_{(1,1)}$} \htext (2.9 .15) {$u_{(2,1)}$} 
\htext (-2.1 -1.2 ) {$\undertilde{Q}$} \htext (-.1 -1.1 ) {$Q$} \htext (1.9 -1.1 ) {$\wtilde{Q}$}}
\caption{The points of the lattice and the corresponding equations.}
\label{fig:3eqs}
\end{figure}
Writing equation (\ref{eq:DEq33point}) explicitly, one arrives at
\begin{eqnarray}
   \left( G_{,u_{(1,0)}}(u_{(1,0)},u_{(0,1)})\,+\,G_{u_{(-1,0)}}(u_{(-1,0)},u_{(0,1)})\right)\,R_{,u_{(-1,0)}} + G(u_{(-1,0)},u_{(0,1)})\,R_{,u_{(-1,0)} u_{(-1,0)}}  \nonumber \\
+\, G(u_{(1,0)},u_{(0,1)}) \left( R_{,u_{(1,0)} u_{(-1,0)}} - \frac{h_{,u_{(1,0)}}(u_{(0,0)},u_{(1,0)})}{h(u_{(0,0)},u_{(1,0)})}\,R_{,u_{(-1,0)}}\right) \label{eq:DEq43point} \\
-\, h(u_{(0,0)},u_{(0,1)}) \left(R_{,u_{(0,0)} u_{(-1,0)}} \,-\,\frac{h_{,u_{(0,0)}}(u_{(0,0)},u_{(1,0)})}{h(u_{(0,0)},u_{(1,0)})}\,R_{,u_{(-1,0)}}\right)\,=\,0\,,  \nonumber
\end{eqnarray}
where we have omitted the arguments of the function $R(n,m,u_{(0,0)},u_{(1,0)},u_{(-1,0)};\alpha,\beta)$ for simplicity, and the equations $Q =  0$, $\undertilde{Q} = 0$ have been taken into account to evaluate the derivatives of $u_{(-1,0)}$ and $u_{(0,0)}$, i.e.
\begin{eqnarray*}
 \frac{\partial u_{(0,0)}}{\partial u_{(1,0)}} = - \frac{Q_{,u_{(1,0)}}}{Q_{,u_{(0,0)}}}  =  - \frac{h(u_{(0,0)},u_{(0,1)})}{G(u_{(1,0)},u_{(0,1)})} \,,& &
\frac{\partial u_{(0,0)}}{\partial u_{(0,1)}} = -\frac{Q_{,u_{(0,1)}}}{Q_{,u_{(0,0)}}}  =  
-\frac{G(u_{(1,0)},u_{(0,1)})}{h(u_{(0,0)},u_{(1,0)})} \,,\\
 \frac{\partial u_{(-1,0)}}{\partial u_{(-1,1)}}= -\frac{\undertilde{Q}_{,u_{(-1,1)}}}{\undertilde{Q}_{,u_{(-1,0)}}}  =  
-\frac{G(u_{(-1,0)},u_{(0,1)})}{h(u_{(-1,1)},u_{(0,1)})} \,, & & 
\frac{\partial u_{(-1,0)}}{\partial u_{(0,0)}} = - \frac{\undertilde{Q}_{,u_{(0,0)}}}{\undertilde{Q}_{,u_{(-1,0)}}} =   -\frac{G(u_{(-1,0)},u_{(0,1)})}{h(u_{(0,0)},u_{(0,1)})} \,.
\end{eqnarray*}
Next, we substitute the derivatives of the polynomials $G$ appearing in Equation (\ref{eq:DEq43point}) by the relation
$$G_{,u_{(1,0)}}(u_{(1,0)},u_{(0,1)}) \,+\,G_{,u_{(-1,0)}}(u_{(-1,0)},u_{(0,1)})\,=\,2\,\frac{G(u_{(1,0)},u_{(0,1)}) - G(u_{(-1,0)},u_{(0,1)})}{u_{(1,0)} - u_{(-1,0)}}\,,$$
which follows from the fact that the polynomial $G$ is quadratic and symmetric in its arguments. 
Upon these substitutions, equation (\ref{eq:DEq43point}) simplifies to
\begin{eqnarray}
 - h(u_{(0,0)},u_{(0,1)}) \left( R_{,u_{(0,0)} u_{(-1,0)}} \,-\,\frac{h_{,u_{(0,0)}}(u_{(0,0)},u_{(1,0)})}{h(u_{(0,0)},u_{(1,0)})}\,R_{,u_{(-1,0)}} \right)   \nonumber \\
 + G(u_{(1,0)},u_{(0,1)}) \left(R_{,u_{(1,0)} u_{(-1,0)}} - \left(\frac{2}{u_{(-1,0)}-u_{(0,0)}}\,+\, \frac{h_{,u_{(1,0)}}(u_{(0,0)},u_{(1,0)})}{h(u_{(0,0)},u_{(1,0)})}\right)\,R_{,u_{(-1,0)}} \right) \label{eq:DEq5a3point} \\
 + G(u_{(-1,0)},u_{(0,1)}) \left( R_{,u_{(-1,0)} u_{(-1,0)}} - \frac{2}{u_{(1,0)} - u_{(-1,0)}} R_{,u_{(-1,0)}}\right) = 0 \,. \nonumber
 \end{eqnarray}
The last equation involves the values of $u$ assigned on four vertices of the lattice (the black ones in Figure \ref{fig:3eqs}) and should hold on every solution of equation $Q=0$. Equation (\ref{eq:DEq5a3point}) depends on $u_{(0,1)}$ through the polynomials $h(u_{(0,0)},u_{(0,1)})$, 
$G(u_{(1,0)},u_{(0,1)})$ and $G(u_{(-1,0)},u_{(0,1)})$, which are in general quadratic in $u_{(0,1)}$. Thus, it is necessary to set the coefficients of different powers of $u_{(0,1)}$ in the latter equation equal to zero, i.e. 
\begin{equation}
\left(\begin{array}{ccc}
h(u_{(0,0)},0) & G(u_{(1,0)},0) & G(u_{(-1,0)},0) \\
h^{\prime}(u_{(0,0)},0) & G^{\prime}(u_{(1,0)},0) & G^{\prime}(u_{(-1,0)},0) \\
h^{\prime\prime}(u_{(0,0)},0) & G^{\prime\prime}(u_{(1,0)},0) & G^{\prime\prime}(u_{(-1,0)},0)
\end{array}\right)\,\left( \begin{array}{l} -R_{,u_{(0,0)} u_{(-1,0)}} + {\mathcal{A}}_1\, R_{,u_{(-1,0)}}\\
 R_{,u_{(1,0)} u_{(-1,0)}} - {\mathcal{A}}_2 \, R_{,u_{(-1,0)}} \\
R_{,u_{(-1,0)} u_{(-1,0)}} - \frac{2 R_{,u_{(-1,0)}}}{u_{(1,0)} - u_{(-1,0)}}
 \end{array}\right)\,=\,\left(\begin{array}{c} 0 \\ 0\\ 0 \end{array}\right) \,, \label{eq:sys3point} \end{equation}
where the prime denotes differentiation w.r.t. $u_{(0,1)}$ and 
$${\mathcal{A}}_1\,=\,\frac{h_{,u_{(0,0)}}(u_{(0,0)},u_{(1,0)})}{h(u_{(0,0)},u_{(1,0)})}\,,\,\,{\mathcal{A}}_2\,=\, \frac{2}{u_{(-1,0)}-u_{(0,0)}}\,+\, \frac{h_{,u_{(1,0)}}(u_{(0,0)},u_{(1,0)})}{h(u_{(0,0)},u_{(1,0)})}\,. $$
In the generic case where
\begin{equation} {\mbox{rank}}\,{\mathcal{G}} \,=\,3\,, \label{defnondegcon} \end{equation}
with
\begin{equation}
{\mathcal{G}}\,=\,\left. \left(\begin{array}{ccc}
h(x,y) & G(x,z) & G(x,w) \\
h_{,x}(x,y) & G_{,x}(x,z) & G_{,x}(x,w) \\
h_{,x x}(x,y) & G_{,x x}(x,z) & G_{,x x}(x,w)\end{array}
 \right)\right|_{x\,=\,0}\,, \label{defnondegcon1}
  \end{equation}
then system (\ref{eq:sys3point}) has the {\sl{unique}} solution
\begin{eqnarray}
R_{,u_{(0,0)} u_{(-1,0)}} &=& \frac{h_{,u_{(0,0)}}(u_{(0,0)},u_{(1,0)})}{h(u_{(0,0)},u_{(1,0)})}\,R_{,u_{(-1,0)}} \,,\nonumber\\
R_{,u_{(1,0)} u_{(-1,0)}} &=& \left(\frac{2}{u_{(-1,0)}-u_{(0,0)}}\,+\, \frac{h_{,u_{(1,0)}}(u_{(0,0)},u_{(1,0)})}{h(u_{(0,0)},u_{(1,0)})}\right)\,R_{,u_{(-1,0)}} \,, \label{eq:detsys}\\
R_{,u_{(-1,0)} u_{(-1,0)}} &=& \frac{2}{u_{(1,0)} - u_{(-1,0)}} R_{,u_{(-1,0)}} \nonumber \,.
\end{eqnarray}
Let it be noted that, even though the rank condition is violated this is a solution of system (\ref{eq:sys3point}) and, consequently, of equation (\ref{eq:DEq5a3point}).

Integrating system (\ref{eq:detsys}), we find that the characteristic reads the form
\begin{equation} 
R(n,m,u_{(0,0)},u_{(1,0)},u_{(-1,0)};\alpha,\beta) \,=\, A(n,m;\alpha,\beta)\, \frac{h(u_{(0,0)},u_{(1,0)})}{u_{(1,0)}-u_{(-1,0)}}\, +\, r(n,m,u_{(0,0)},u_{(1,0)};\alpha,\beta) \,,
\label{eq:Ra3point} \end{equation}
where $A$ and $r$ are arbitrary functions of their arguments.

Next, we substitute (\ref{eq:Ra3point}) into the determining equation (\ref{eq:DEq13point}). Using the relation\footnote{To prove this relation, we write 
$h(u_{(0,0)},u_{(1,0)}) = p_2(u_{(0,0)}) u_{(1,0)}^2 + p_1(u_{(0,0)}) u_{(1,0)} + p_0(u_{(0,0)})$ and $h(u_{(-1,0)},u_{(0,0)}) = {\mathcal{S}}_n^{-1}(h(u_{(0,0)},u_{(1,0)}))$, since the latter function is the backward shift of the former in the $n$-direction. Using the fact that they are symmetric, a straightforward calculation implies the result.}
\begin{equation}  h_{,u_{(1,0)}}(u_{(0,0)},u_{(1,0)})\,+\,h_{,u_{(-1,0)}}(u_{(-1,0)},u_{(0,0)})\,=\,2\,\frac{h(u_{(0,0)},u_{(1,0)})\,-\,h(u_{(-1,0)},u_{(0,0)})}{u_{(1,0)} - u_{(-1,0)}}\,,\label{hrel1} \end{equation}
and its shifted versions, we find that it is convenient to set
$$r(n,m,u_{(0,0)},u_{(1,0)};\alpha,\beta) = \frac{-A(n,m;\alpha,\beta) h_{,u_{(1,0)}}(u_{(0,0)},u_{(1,0)}) + 
\phi(n,m,u_{(0,0)},u_{(1,0)};\alpha,\beta)}{2}\,.$$ 
Upon these substitutions the characteristic takes the form
\begin{equation} 
 A(n,m;\alpha,\beta)\,\left( \frac{h(u_{(0,0)},u_{(1,0)})}{u_{(1,0)}-u_{(-1,0)}}\,-\,\frac{1}{2}h_{,u_{(1,0)}}(u_{(0,0)},u_{(1,0)})\right)\, +\, \frac{1}{2} \phi(n,m,u_{(0,0)},u_{(1,0)};\alpha,\beta) 
\end{equation}
and the determining equation becomes
\begin{eqnarray}
 Q_{,u_{(0,0)}} \left(A(n,m;\alpha,\beta) \left(-h_{,u_{(1,0)}}(u_{(0,0)},u_{(1,0)}) + \frac{2 h(u_{(0,0)},u_{(1,0)})}{u_{(1,0)}-u_{(-1,0)}} \right) + 
\phi \right) & \nonumber \\
& \nonumber  \\
 + Q_{,u_{(1,0)}} \left(A(n+1,m;\alpha,\beta) \left(-h_{,u_{(2,0)}}(u_{(1,0)},u_{(2,0)})+\frac{2h(u_{(1,0)},u_{(2,0)})}{u_{(2,0)}-u_{(0,0)}} \right) + 
{\mathcal{S}}_n (\phi) \right)& \nonumber  \\
& \label{eq:DEq63point} \\
 + Q_{,u_{(0,1)}} \left(A(n,m+1;\alpha,\beta) \left(-h_{,u_{(1,1)}}(u_{(0,1)},u_{(1,1)})+\frac{2 h(u_{(0,1)},u_{(1,1)})}{u_{(1,1)}-u_{(-1,1)}} \right) + 
{\mathcal{S}}_m (\phi) \right) & \nonumber  \\
& \nonumber  \\
 + Q_{,u_{(1,1)}} \left( A(n+1,m+1;\alpha,\beta) \left(-h_{,u_{(2,1)}}(u_{(1,1)},u_{(2,1)})+\frac{2 h(u_{(1,1)},u_{(2,1)})}{u_{(2,1)}-u_{(0,1)}} \right) 
+  {\mathcal{S}}_n ({\mathcal{S}}_m (\phi))  \right)= 0,& \nonumber 
\end{eqnarray}
where we have omitted the arguments of the function $\phi$ and its shifted values. Equation (\ref{eq:DEq63point}) involves the values of the function $u$ assigned on the eight vertices of Figure \ref{fig:3eqs}. Using the equations $Q=0$ on the three faces, we eliminate three of these values, and we have chosen to eliminate $u_{(-1,0)}$, $u_{(1,1)}$ and $u_{(2,1)}$. 

The only terms in (\ref{eq:DEq63point}) which depend on $u_{(-1,0)}$ and $u_{(-1,1)}$ appear in the coefficients of $A(n,m;\alpha,\beta)$ and $A(n,m+1;\alpha,\beta)$, respectively. 
We take the total derivative of the determining equation (\ref{eq:DEq63point}) w.r.t. $u_{(-1,1)}$, since $u_{(-1,0)}$ depends on $u_{(-1,1)}$ through the equation $\undertilde{Q}=0$. After a lengthy calculation, this simplifies to 
\begin{equation} 
\left( A(n,m;\alpha,\beta)- A(n,m+1;\alpha,\beta) \right) \,h(u_{(0,0)},u_{(1,0)})\,h(u_{(0,0)},u_{(0,1)})\,G(u_{(1,0)},u_{(0,1)})\,=\,0\,, \label{eq:adef1}
\end{equation}
where we have used equations $Q=0$, $\undertilde{Q}=0$ to eliminate $u_{(1,1)}$ and $u_{(-1,0)}$, respectively, and relation (\ref{rel2}).  Equation (\ref{eq:adef1}) implies that $A(n,m;\alpha,\beta)$ does not depend on $m$, i.e.
\begin{equation} A(n,m;\alpha,\beta) = a(n;\alpha,\beta) \,.\label{eq:adef} \end{equation}
Thus, taking into account the latter, the terms in (\ref{eq:DEq63point}) involving $A(n,m;\alpha,\beta)$ and $A(n,m+1;\alpha,\beta)$ are independent of $u_{(-1,1)}$, and consequently they can be evaluated at $u_{(-1,1)}=0$, simplifying to
\begin{eqnarray} 
 a(n;\alpha,\beta) \left [ Q_{,u_{(0,0)}}\left(-h_{,u_{(1,0)}}(u_{(0,0)},u_{(1,0)}) + 2 h(u_{(0,0)},u_{(1,0)}) 
\frac{Q_{,u_{(1,0)}}(u_{(0,0)},u_{(1,0)},u_{(0,1)},0)}{Q(u_{(0,0)},u_{(1,0)},u_{(0,1)},0)}  \right)\right. \nonumber \\ 
\left. + Q_{,u_{(0,1)}} \left( \frac{2 h(u_{(0,1)},u_{(1,1)})}{u_{(1,1)}} -  h_{,u_{(1,1)}}(u_{(0,1)},u_{(1,1)}) \right) \right ] \,. \label{eq:DEq8a3point}
\end{eqnarray}

Moreover, it turns out that the determining equation, apart from the function $\phi$, does not depend on the value $u_{(2,0)}$. Indeed, substituting (\ref{eq:adef}) in equation (\ref{eq:DEq63point}), the total derivative of the resulting equation w.r.t. $u_{(2,0)}$ is identically zero. Thus, the relevant terms in (\ref{eq:DEq63point}), except the function $\phi$,  are independent of $u_{(2,0)}$, and hence we can set $u_{(2,0)}=0$ in them. These can be written as  
\begin{subequations}
\begin{eqnarray}
 \left . \left( -h_{,u_{(2,0)}}(u_{(1,0)},u_{(2,0)}) + \frac{2 h(u_{(1,0)},u_{(2,0)})}{u_{(2,0)}-u_{(0,0)}} \right) \right |_{u_{(2,0)} = 0} & = & u_{(0,0)}^2 \partial_{u_{(0,0)}} \left(\frac{h(u_{(0,0)},u_{(1,0)})}{u_{(0,0)}^2}\right)  \,, \label{eq:DEq7a13point} \\
 \left . \left( -h_{,u_{(2,1)}}(u_{(1,1)},u_{(2,1)}) + \frac{2 h(u_{(1,1)},u_{(2,1)})}{u_{(2,1)}-u_{(0,1)}} \right) \right |_{u_{(2,0)} = 0} & = &h_{,u_{(0,1)}}(u_{(0,1)},u_{(1,1)})  \nonumber \\
& & - 2 h(u_{(0,1)},u_{(1,1)}) \left.  \frac{Q_{,u_{(0,1)}}}{Q} \right |_{u_{(0,0)}=0} \,.\label{eq:DEq7a23point} 
\end{eqnarray}
\label{eq:DEq7aa3point}
\end{subequations}

Substituting (\ref{eq:adef}), (\ref{eq:DEq8a3point}) and (\ref{eq:DEq7aa3point}) in equation (\ref{eq:DEq63point}), the latter simplifies to
\begin{eqnarray}
 a(n;\alpha,\beta) \left[Q_{,u_{(0,0)}}\left(-h_{,u_{(1,0)}}(u_{(0,0)},u_{(1,0)}) + 2 h(u_{(0,0)},u_{(1,0)}) 
\frac{Q_{,u_{(1,0)}}(u_{(0,0)},u_{(1,0)},u_{(0,1)},0)}{Q(u_{(0,0)},u_{(1,0)},u_{(0,1)},0)}  \right) \right. \nonumber \\ 
 \nonumber \\ 
  \left.  + Q_{,u_{(0,1)}} \left( \frac{2 h(u_{(0,1)},u_{(1,1)}) }{u_{(1,1)}} -  h_{,u_{(1,1)}}(u_{(0,1)},u_{(1,1)}) \right) \right] \nonumber \\
  \nonumber\\
 + a(n+1;\alpha,\beta) \left[ Q_{,u_{(1,0)}} \frac{u_{(0,0)} h_{,u_{(0,0)}}(u_{(0,0)},u_{(1,0)}) - 2 h(u_{(0,0)},u_{(1,0)})}{u_{(0,0)}}\right. \nonumber \\
 \nonumber \\
  \left.  + Q_{,u_{(1,1)}} \left( h_{,u_{(0,1)}}(u_{(0,1)},u_{(1,1)}) - 2 h(u_{(0,1)},u_{(1,1)}) \frac{Q_{,u_{(0,1)}}(0,u_{(1,0)},u_{(0,1)},u_{(1,1)})}{Q(0,u_{(1,0)},u_{(0,1)},u_{(1,1)})}\right)  \right] \label{eq:DEq93point} \\
 \nonumber \\
  + Q_{,u_{(0,0)}} \phi(n,m,u_{(0,0)},u_{(1,0)};\alpha,\beta) + Q_{,u_{(1,0)}} \phi(n+1,m,u_{(1,0)},u_{(2,0)};\alpha,\beta) \nonumber \\
 \nonumber \\ 
  + Q_{,u_{(0,1)}} \phi(n,m+1,u_{(0,1)},u_{(1,1)};\alpha,\beta)+ Q_{,u_{(1,1)}} \phi(n+1,m+1,u_{(1,1)},u_{(2,1)};\alpha,\beta) = 0\,. \nonumber
\end{eqnarray}

It remains to simplify the coefficients of $a(n;\alpha,\beta)$ and $a(n+1;\alpha,\beta)$ by eliminating the value $u_{(1,1)}$.
For this purpose, we use the relations (\ref{rel3}),(\ref{rel4}) and the following ones
\begin{eqnarray}
 \frac{u_{(1,1)} \left(h_{,u_{(1,1)}}(u_{(0,1)},u_{(1,1)}) + G_{,u_{(1,0)}}(u_{(1,0)},u_{(0,1)}) \right)- 2 h(u_{(0,1)},u_{(1,1)})}{2 u_{(1,1)} 
G(u_{(1,0)},u_{(0,1)})}= \left. \frac{Q_{,u_{(1,0)}}}{Q} \right |_{u_{(1,1)}=0}\,, & & \\ 
& & \nonumber \\
 \frac{u_{(0,0)} \left( h_{,u_{(0,0)}}(u_{(0,0)},u_{(1,0)}) + G_{,u_{(0,1)}}(u_{(1,0)},u_{(0,1)}) \right) -2 h(u_{(0,0)},u_{(1,0)})}{2 u_{(0,0)} 
G(u_{(1,0)},u_{(0,1)})}  = \left. \frac{Q_{,u_{(0,1)}}}{Q} \right|_{u_{(0,0)}=0} \,,& & \\
& & \nonumber \\
 \frac{h(u_{(0,0)},u_{(1,0)}) G_{,u_{(1,0)}}(u_{(1,0)},u_{(0,1)}) -h_{,u_{(1,0)}}(u_{(0,0)},u_{(1,0)}) 
G(u_{(1,0)},u_{(0,1)})}{h(u_{(0,1)},u_{(1,1)}) G_{,u_{(0,1)}}(u_{(1,0)},u_{(0,1)}) -h_{,u_{(0,1)}}(u_{(0,1)},u_{(1,1)}) G(u_{(1,0)},u_{(0,1)})} = 
\frac{Q_{u_{(1,1)}}}{Q_{u_{(0,0)}}}\,, && 
\end{eqnarray}
which hold in view of the equation $Q=0$. 

Finally, equation (\ref{eq:DEq93point}) simplifies to
\begin{eqnarray}
 \left( a(n;\alpha,\beta) - a(n+1;\alpha,\beta) \right) \,h(u_{(0,0)},u_{(1,0)})^2\,\partial_{u_{(1,0)}} \left( \frac{G(u_{(1,0)},u_{(0,1)})}{h(u_{(0,0)},u_{(1,0)})} \right) &&\nonumber\\
&& \nonumber \\
 + G(u_{(1,0)},u_{(0,1)}) \phi(n,m,u_{(0,0)},u_{(1,0)};\alpha,\beta)  + h(u_{(0,0)},u_{(0,1)}) \phi(n+1,m,u_{(1,0)},u_{(2,0)};\alpha,\beta) && \label{eq:DEq103point} \\
&& \nonumber \\
 + h(u_{(0,0)},u_{(1,0)}) \phi(n,m+1,u_{(0,1)},u_{(1,1)};\alpha,\beta) = Q_{,u_{(1,1)}}^2 \phi(n+1,m+1,u_{(1,1)},u_{(2,1)};\alpha,\beta)  \,.&& \nonumber
 \end{eqnarray}

Obviously, if we take $\phi(n,m,u_{(0,0)},u_{(1,0)};\alpha,\beta)=0$ and $a(n;\alpha,\beta)=constant$, then the last equation is satisfied. Thus, so far, we have proved that, independently of the rank condition (\ref{defnondegcon}), every equation $Q=0$, where the function $Q$ is affine linear and possesses the ${\mathrm{D}}_4$-symmetry, admits a three-point symmetry with characteristic 
$$R(n,m,u_{(0,0)},u_{(1,0)},u_{(-1,0)};\alpha,\beta) = \frac{h(u_{(0,0)},u_{(1,0)};\alpha,\beta)}{u_{(1,0)}-u_{(-1,0)}} - \frac{1}{2} h_{,u_{(1,0)}}(u_{(0,0)},u_{(1,0)};\alpha,\beta) \,. $$

Since the function $Q$ is symmetric, i.e. 
$$Q(u_{(0,0)},u_{(1,0)},u_{(0,1)},_{(1,1)};\alpha,\beta) \,= \,\epsilon\, Q(u_{(0,0)},u_{(0,1)},u_{(1,0)},_{(1,1)};\beta,\alpha)\,,$$
we find that the characteristic of another symmetry generator is
$$R(n,m,u_{(0,0)},u_{(0,1)},u_{(0,-1)};\beta,\alpha) = \frac{h(u_{(0,0)},u_{(0,1)};\beta,\alpha)}{u_{(0,1)}-u_{(0,-1)}} - \frac{1}{2} h_{,u_{(0,1)}}(u_{(0,0)},u_{(0,1)};\beta,\alpha) \,. $$

Now, we focus in the generic case, i.e. ${\mbox{rank}}\,{\mathcal{G}} = 3$. In this case, the characteristic of a three-point generalized symmetry generator is necessarily of the form
\begin{eqnarray*}
 R(n,m,u_{(0,0)},u_{(1,0)},u_{(-1,0)},\alpha,\beta)& =& a(n;\alpha,\beta) \left(\frac{h(u_{(0,0)},u_{(1,0)};\alpha,\beta)}{u_{(1,0)}-u_{(-1,0)}} - \frac{1}{2} h_{,u_{(1,0)}}(u_{(0,0)},u_{(1,0)};\alpha,\beta)\right) \\
 & &  +\, \frac{1}{2}\,\phi(n,m,u_{(0,0)},u_{(1,0)};\alpha,\beta)\,,
\end{eqnarray*}
where the functions $a(n,\alpha,\beta)$, $\phi(n,m,u_{(0,0)},u_{(1,0)};\alpha,\beta)$ satisfy the determining equation (\ref{eq:DEq103point}). 

Actually, the function $\phi(n,m,u_{(0,0)},u_{(1,0)};\alpha,\beta)$ should be independent of its fourth argument, namely $u_{(1,0)}$. Indeed, differentiating the determining equation (\ref{eq:DEq103point}) w.r.t. $u_{(2,0)}$ and using (\ref{rel2}), we arrive at
\begin{eqnarray} 
h(u_{(0,1)},u_{(1,1)})\, \phi_{,u_{(2,0)}}(n+1,m,u_{(1,0)},u_{(2,0)};\alpha,\beta)\, + \nonumber \\
G(u_{(1,0)},u_{(0,1)})\, \phi_{,u_{(2,1)}}(n+1,m+1,u_{(1,1)},u_{(2,1)};\alpha,\beta)\, \frac{\partial u_{(2,1)}}{\partial u_{(2,0)}}\,=\,0\,, \label{eq:DEq113point} 
\end{eqnarray}
since $u_{(2,1)}$ depends on $u_{(2,0)}$ through the equation $\wtilde{Q}=0$. 
Since the value $u_{(0,1)}$ occurs in the last equation through the polynomials $h(u_{(0,1)},u_{(1,1)})$ and $G(u_{(1,0)},u_{(0,1)})$, we set the coefficients of the different powers of $u_{(0,1)}$ equal to
zero. This leads to a linear system for the derivatives of $\phi$ appearing in equation (\ref{eq:DEq113point}). 
The maximal rank condition for the matrix ${\mathcal{G}}$ ensures that the matrix of this system given by
$$\left. \left(\begin{array}{cc} 
h(u_{(0,1)},u_{(1,1)}) & G(u_{(1,0)},u_{(0,1)}) \\
h^{\prime}(u_{(0,1)},u_{(1,1)}) & G^{\prime}(u_{(1,0)},u_{(0,1)}) \\
h^{\prime \prime}(u_{(0,1)},u_{(1,1)}) & G^{\prime \prime}(u_{(1,0)},u_{(0,1)})
\end{array} 
 \right)\right|_{u_{(0,1)}=0} \,,\qquad {\mbox{where}}\quad{}^{\prime} \,\equiv\,\frac{\partial {\phantom {u_{(1,0)}}}}{\partial u_{(0,1)}}\,,$$
has rank 2, which implies that the function $\phi$ is independent of its fourth argument, i.e.
$$\phi(n,m,u_{(0,0)},u_{(1,0)};\alpha,\beta)\,=\,\phi(n,m,u_{(0,0)};\alpha,\beta)\,. $$

Thus, in this case, the determining equation becomes
\begin{eqnarray}
\left( a(n;\alpha,\beta) - a(n+1;\alpha,\beta) \right) \,h(u_{(0,0)},u_{(1,0)})^2\,\partial_{u_{(1,0)}} \left( \frac{G(u_{(1,0)},u_{(0,1)})}{h(u_{(0,0)},u_{(1,0)})} \right) && \nonumber \\
&& \nonumber \\
+ G(u_{(1,0)},u_{(0,1)}) \phi(n,m,u_{(0,0)};\alpha,\beta)  + h(u_{(0,0)},u_{(0,1)}) \phi(n+1,m,u_{(1,0)};\alpha,\beta) && \label{eq:fineq13pap}\\
&& \nonumber \\
+ h(u_{(0,0)},u_{(1,0)}) \phi(n,m+1,u_{(0,1)};\alpha,\beta) = Q_{,u_{(1,1)}}^2 \phi(n+1,m+1,u_{(1,1)};\alpha,\beta)  \,.&&\nonumber  
\end{eqnarray}

A final comment is that the form of the function $\phi$ is obtained in a similar manner as the one used to obtain the
general form of the characteristic of a Lie point symmetry generator, as presented in Section \ref{lie}. The substitution of $\phi$ into the determining equation (\ref{eq:fineq13pap}) and the usage of equation $Q=0$ to eliminate $u_{(1,1)}$ in the resulting equation yield a polynomial in $u_{(0,0)}$, $u_{(1,0)}$ and $u_{(0,1)}$.
Setting the coefficients of the different monomials equal to zero, we come up with an overdetermined linear system of difference equations for the unknown function $a(n;\alpha,\beta)$ and the functions $A_i(n,m;\alpha,\beta)$,
which occur in the general form of the function $\phi$. The general solution of this system delivers the three-point generalized symmetries and the point symmetries, as well.

\end{document}